\documentclass[%
reprint,
superscriptaddress,
amsmath,amssymb,
aps,
prb,
]{revtex4-2}

\usepackage{graphicx}
\usepackage{dcolumn}
\usepackage{bm}
\usepackage{dsfont}
\usepackage{braket}
\usepackage{hyperref}
\usepackage{xcolor,soul}
\usepackage{multirow} 
\usepackage{changes}

\begin{document}


\title{Influence of temperature, doping, and amorphization on the electronic structure and magnetic damping of iron}

\author{Zhihao Jiang}
\affiliation{Department of Materials Science and Engineering, University of Illinois, Urbana-Champaign, Urbana, IL 61801, USA}
\author{Axel Hoffmann}
\affiliation{Department of Materials Science and Engineering, University of Illinois, Urbana-Champaign, Urbana, IL 61801, USA}
\affiliation{Materials Research Laboratory, University of Illinois at Urbana-Champaign, Urbana, IL 61801, USA}
\author{André Schleife}%
\email{schleife@illinois.edu}
\affiliation{Department of Materials Science and Engineering, University of Illinois, Urbana-Champaign, Urbana, IL 61801, USA}
\affiliation{Materials Research Laboratory, University of Illinois at Urbana-Champaign, Urbana, IL 61801, USA}
\affiliation{National Center for Supercomputing Applications, University of Illinois at Urbana-Champaign, Urbana, IL 61801, USA}

\date{\today}

\begin{abstract}
Hybrid magnonic quantum systems have drawn increased attention in recent years for coherent quantum information processing, but too large magnetic damping is a persistent concern when metallic magnets are used.
Their intrinsic damping is largely determined by electron-magnon scattering induced by spin-orbit interactions. 
In the low scattering limit, damping is dominated by intra-band electronic transitions, which has been theoretically shown to be proportional to the electronic density of states at the Fermi level.
In this work, we focus on body-centered-cubic iron as a paradigmatic ferromagnetic material.
We comprehensively study its electronic structure using first-principles density functional theory simulations and account for finite lattice temperature, boron (B) doping, and structure amorphization.
Our results indicate that temperature induced atomic disorder and amorphous atomic geometries only have a minor influence.
Instead, boron doping noticeably decreases the density of states near the Fermi level with an optimal doping level of 6.25\,\%.
In addition, we show that this reduction varies significantly for different atomic geometries and report that the highest reduction correlates with a large magnetization of the material.
This may suggest materials growth under external magnetic fields as a route to explore in experiment. 
\end{abstract}

\maketitle

\let\vec\mathbf


\section{\label{sec1:introduction}Introduction}
Hybrid magnonics is gaining growing interest due to its potential for coherent quantum information processing \cite{Li2020,Awschalom2021,Yuan2022,Jiang2023}.
This was triggered by the experimental demonstration of coherent coupling between a magnon and a superconducting qubit, mediated by a microwave cavity, by Tabuchi \emph{et al.}\ \cite{Tabuchi2015}
As one key ingredient in hybrid magnonic systems, magnons possess unique advantages such as easily tunable resonance frequencies through external magnetic fields or materials anisotropy, microwave bandwidths that match state-of-the-art superconducting quantum devices, and intrinsic non-reciprocity that is promising for noise-resilient quantum state transduction~\cite{Rezende2020, Jiang2023}.
In recent years, more research has established coupling between magnons and microwave photons in a cavity or a coplanar circuit structure \cite{Kostylev2016, Bhoi2014, Li2019PRL, Hou2019}.

However, one essential challenge is the damping of magnon excitations that limits the coherence time of hybrid quantum states \cite{Jiang2023,Schmidt2020}. 
Therefore, exploring materials with low magnetic damping is crucial for achieving hybrid magnonic quantum devices with long coherence time.
One of the best-known materials in this context is the ferrimagnetic insulator yttrium iron garnet (YIG), Y$_3$Fe$_5$O$_{12}$, whose magnetic damping parameter can be as low as $10^{-5}$ in bulk crystals \cite{Glass1976}.
However, insulators such as YIG are not desired for many spintronics applications that require a charge current through the material~\cite{Hirohata2020}.
In addition, YIG is not well suited to be integrated into on-chip devices for circuit quantum electrodynamics due to experimental constraints. 
One of the reasons, is that in order to have sufficient crystalline quality for low damping, YIG films need to be grown on specific substrates for achieving epitaxy.  
The standard substrate choice for this task is gadolinium gallium garnet (GGG, Gd$_3$Ga$_5$O$_{12}$), but at very low temperatures the increased magnetic susceptibility due to ordering of the Gd magnetic moments can increase the magnetic damping of YIG films considerably \cite{Jermain2017,Danilov1989,Danilov2002,Mihalcenu2018,Kosen2019}.
Ferromagnetic metals and alloys are an alternative category of promising materials that are significantly easier to integrate on-chip.
It is therefore desirable to explore and optimize metallic magnets towards low magnetic damping for hybrid magnonic quantum devices~\cite{Azzawi2023}.

A major contribution to magnetic damping in \emph{metallic} magnets arises from conduction electrons that dissipate magnons through the spin-orbit (SO) interaction \cite{Korenman1972, Kambersky1976, Gilmore2007}.
This mechanism is described by two early theories, the breathing Fermi surface (BFS) model and the later torque correlation (TC) model that were developed by Kambersk{\'y} \cite{Kambersky1970,Kambersky1976}.
The TC model is more general and describes how the SO torque $\braket{n, \vec k|[\vec{\sigma}^-, \hat{H}_\text{SO}]|m, \vec k}$ induces intra-band ($m=n$) and inter-band ($m \neq n$) electronic transitions which are interpreted as conductivity-like and resistivity-like damping pathways for magnons, respectively \cite{Kambersky2007, Gilmore2007, Gilmore2008, Thonig2014}.
In its low scattering limit, when the spectral overlap between different bands is small, damping is largely dominated by the intra-band part, in agreement with the BFS model.
Damping is shown to be approximately proportional to the electronic density of states (EDOS) at the Fermi level~\cite{Kambersky1970, Kambersky1976, Mizukami2011}.
Many experimental works confirm that the lowest damping coincides with the lowest EDOS at the Fermi level (EDOS-FL) for ferromagnetic alloys such as cobalt-iron (Co-Fe) and iron-vanadium (Fe-V) \cite{Schoen2016,Schoen2017,Arora2021}.
Therefore, one practical way to reduce magnetic damping is to minimize EDOS-FL by electronic-structure engineering.

In this work, we focus on the paradigmatic ferromagnetic material, body-centered-cubic (BCC) iron (Fe).
We use first-principles electronic-structure theory to comprehensively study its EDOS-FL under the conditions of (i) thermal atomic disorder, (ii) boron (B) doping, and (iii) structure amorphization.
We account for the effect of thermal disorder on EDOS-FL since real devices always work at non-zero temperature, although for superconducting quantum devices the operating temperature can be extremely low.
Typical superconducting quantum circuits, such as qubits, are operated at temperatures of a few 10s mK.
Meanwhile, recent experimental work has shown that doping carbon (C) or B into ferromagnetic alloys can make the structure amorphous and reduce magnetic damping \cite{Wang2019,Lourembam2021}.
Hence, it is interesting to investigate the origin of the reduced damping in doped amorphous alloys and the extent to which it can be attributed to reduced EDOS-FL.
To explore this, we study the effects of B doping and structure amorphization individually and compare EDOS-FL for \textit{crystalline} Fe doped with B and \textit{amorphous} Fe, as well as \textit{pure crystalline} Fe.
Finally, we consider amorphous Fe with B doping to investigate the combined effect.

The remainder of this paper is organized as follows:
We first introduce the methods that are used to construct our simulation cells including BCC Fe supercells with phonon excitations, BCC Fe supercells doped with B, amorphous Fe supercells, and amorphous Fe supercells doped with B, in Sec.~\ref{sec2:methods}.
The details of our DFT simulations are also explained in Sec.~\ref{sec2:methods}.
In Sec.~\ref{sec3:results}, we discuss our main results for the EDOS of the different structures.
In general, we find that phonon excitations do not significantly change EDOS-FL up to at least 500 K.
Doping B into BCC Fe decreases EDOS-FL, while making the Fe amorphous increases EDOS-FL.
Overall, amorphous Fe doped with B can have lower EDOS-FL than pure BCC Fe.
Finally, we give our conclusions in Sec.~\ref{sec4:conclusion}.

\section{\label{sec2:methods}Methodology}

\begin{figure}
\includegraphics[width=0.95\columnwidth]{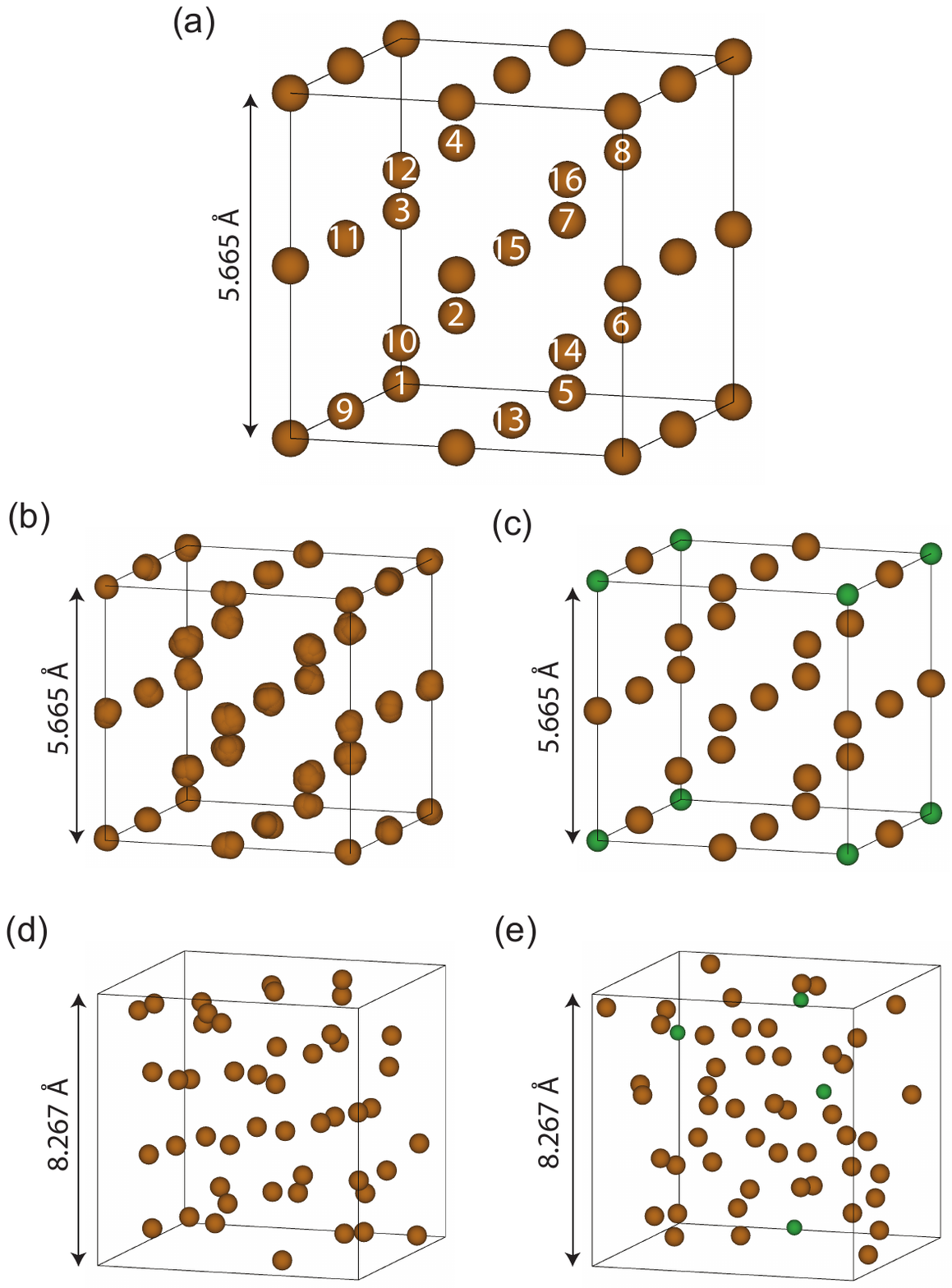}
\caption{\label{fig-1:structures} 
Simulation supercells used in this work.
(a) The $2 \times 2 \times 2$ supercell of ideal BCC Fe at the temperature of $T=0$ K.
The supercell includes 16 Fe atoms (brown spheres) that are labeled by the numbers on them.
(b) BCC Fe supercell at $T=300$ K, with 50 instantaneous snapshots of the vibrating lattice superimposed.
(c) BCC Fe supercell doped with 6.25\% B (green spheres), corresponding to $n_\mathrm{B}=1$ in Table \ref{table1:cluster expansion}.
(d) Snapshot of amorphous Fe (a-Fe) with a volume of a cubic $3 \times 3 \times 3$ supercell.
This supercell is cut from a larger one constructed by molecular dynamics (MD) simulations (see text).
(e) Snapshot of amorphous Fe doped with 7.5\% B (a-FeB). 
All these supercell images are produced by VESTA~\cite{Momma2011}.
}
\end{figure}

Thermal disorder of the atomic positions in the BCC Fe lattice is modeled by considering phonons that displace nuclei from their ideal 0 K lattice positions of the BCC crystal structure [see Fig.\ \ref{fig-1:structures}(a)].
Simulation cells of the thermally disordered lattice are constructed by superimposing harmonic phonon modes with random phases and amplitudes according to classical statistics at different temperatures \cite{Aberg2013}.
In this work, we consider three different temperatures of 10 K, 300 K, and 500 K.
For each temperature, 50 random snapshots of the disordered $2 \times 2 \times 2$ BCC Fe supercell are generated for simulations.
In Fig.\ \ref{fig-1:structures}(b), we show the disordered supercell superimposing 50 snapshots for $T=300$~K as an example.
The supercell snapshots for $T=10$~K and $T=500$~K are similar but with different amplitudes of the atomic displacements.
For analyzing the EDOS, we average over all 50 snapshots for each temperature.  

\begin{table}
\begin{tabular}{|c|c|c|c|} 
\hline
$n_\text{B}$ & class & degeneracy & representation \\ \hline
1 & 1 & 16 & 1 \\
\hline
\multirow{4}{*}{2} & 2 & 64 & 1, 2\\ 
                    & 3 & 24 & 1, 3\\ 
                    & 4 & 24 & 1, 7\\
                    & 5 & 8 & 1, 15\\ \hline
\multirow{6}{*}{3} & 6 & 192 & 1, 2, 3\\ 
                    & 7 & 192 & 1, 2, 7\\ 
                    & 8 & 64 & 1, 2, 15\\
                    & 9 & 48 & 1, 3, 5\\
                    & 10 & 48 & 1, 3, 13\\
                    & 11 & 16 & 1, 7, 11\\ \hline
\multirow{18}{*}{4} & 12 & 48 & 1, 2, 3, 4\\ 
                         & 13 & 384 & 1, 2, 3, 5\\ 
                         & 14 & 96 & 1, 2, 3, 6\\
                         & 15 & 192 & 1, 2, 3, 8\\
                         & 16 & 384 & 1, 2, 3, 13\\
                         & 17 & 96 & 1, 2, 3, 14\\
                         & 18 & 96 & 1, 2, 3, 16\\ 
                         & 19 & 48 & 1, 2, 7, 8\\
                         & 20 & 128 & 1, 2, 7, 11\\
                         & 21 & 96 & 1, 2, 7, 12\\
                         & 22 & 96 & 1, 2, 7, 16\\
                         & 23 & 16 & 1, 2, 15, 16\\ 
                         & 24 & 12 & 1, 3, 5, 7\\
                         & 25 & 16 & 1, 3, 5, 9\\
                         & 26 & 48 & 1, 3, 5, 11\\
                         & 27 & 48 & 1, 3, 5, 15\\
                         & 28 & 12 & 1, 3, 13, 15\\ 
                         & 29 & 4 & 1, 7, 11, 13\\ \hline
\end{tabular}
\caption{\label{table1:cluster expansion}
All non-equivalent classes (second column) of arranging $n_\text{B}$ B atoms (first column) on 16 Fe BCC lattice sites, Fe$_{16-n_\text{B}}$B$_{n_\text{B}}$.
Only these representative structures are simulated in our work.
The third column provides the degeneracy for each class determined by crystal structure symmetry and the fourth column contains one representative atomic geometry that was used in this work (see atom labels in Fig.\ \ref{fig-1:structures}).
}
\end{table}

Lattice geometries for B doped BCC Fe are constructed using a cluster-expansion method which was used before for studying binary alloys \cite{Schleife2010, Schleife2011, Sanchez1984, Zunger1994, Teles2000, Caetano2006}.
This method starts with a $2 \times 2 \times 2$ BCC Fe supercell with 16 atoms and replaces $n_\mathrm{B}$ Fe atoms with B atoms.
Generally, there are $C_{n_\mathrm{Fe}+n_\mathrm{B}}^{n_\mathrm{B}}$ ways of distributing $n_\mathrm{Fe}$ Fe atoms and $n_\mathrm{B}$ B atoms, however, symmetries of the lattice structure reduce the total number of non-equivalent configurations to a few classes with different folds of degeneracy (see Table~\ref{table1:cluster expansion} for all different $n_\text{B}$ considered in this paper).
Electronic-structure simulations are carried out for only one representative of each class, which significantly reduces the computational effort and allows us to study an alloy with $n_\mathrm{Fe}$+$n_\mathrm{B}$=16 atoms.
In this work, we consider $n_\mathrm{B}$=1 (see Fig.~\ref{fig-1:structures}(c) for Fe$_{15}$B$_{1}$), 2, 3, and 4 that correspondingly have 1, 4, 6, and 18 non-equivalent classes (see Table \ref{table1:cluster expansion}).
For each doping level $n_\mathrm{B}$, we compute the EDOS by averaging over the non-equivalent classes weighted by their degeneracies (third column in Table \ref{table1:cluster expansion}).
There are different methods to determine the weights used for averaging, introduced in Ref.\ \cite{Schleife2010}, corresponding to different thermodynamic conditions of the alloy.
The weights we use here correspond to a simplification of the strict regular solution, with the macroscopic alloy composition or doping level consisting exclusively of microscopic structures with exactly the same concentration of elements.

We construct amorphous Fe (a-Fe) atomic geometries by simulating heating and quenching processes using molecular dynamics (MD) simulations as implemented in the LAMMPS code~\cite{Thompson2022}.
The inter-atomic interactions between Fe atoms are described via a `magnetic' interatomic potential that was developed by Dudarev and Derlet \cite{Dudarev2005, Dudarev2007correction, Derlet2007}, based on the embedded atom method.
The inter-atomic potential for our simulation is obtained from the OpenKim website \cite{OpenKim1, OpenKim2, OpenKim3, OpenKim4}.
The same potential has been used previously \cite{Ma2007} to generate a-Fe geometries and was shown to produce characteristic features observed in the experimentally measured radial distribution function (RDF) $g(r)$~\cite{Ichikawa1973}.
In our NVT canonical ensemble MD simulations, we use a $10 \times 10 \times 10$ supercell of BCC Fe with 2000 atoms.
The system is then heated to 10,000 K using a Nos{\'e}-Hoover thermostat, leading to a completely liquid state.
We then implement a cooling process where first the temperature $T$ is decreased to 4,000~K in 50~picoseconds (ps) and maintained at 4,000~K for 250~ps.
Subsequently, $T$ is decreased to 3,000~K in 250~ps.
Finally, $T$ is decreased to 300 K (room temperature) in 50~ps and maintained for another 50~ps.
We compute the RDF $g(r)$ for this final structure and find good agreement with the result from Ma \emph{et al.}\ \cite{Ma2007} and the experimental observation from Ichikawa~\cite{Ichikawa1973} (see Fig.~1 in the supplemental information).

We then extract smaller cubic cells from this result, corresponding to volumes of a $3 \times 3 \times 3$ BCC Fe cell [see Fig.\ \ref{fig-1:structures}(d)], and use these for first-principles calculations of the electronic structure.
We cut 10 such a-Fe snapshots from random positions of the large a-Fe structure, simulate all of them, and average the results.
To generate a-Fe doped with B (a-FeB), we use a single a-Fe snapshot and randomly replace Fe atoms by B atoms according to the doping level [see Fig.\ \ref{fig-1:structures}(e)].
For each doping level, 10 different configurations of random replacement are constructed for our simulations in order to approach a statistical average.

For all these different atomic geometries, we compute electronic densities of states (EDOS) using density functional theory (DFT) as implemented in the Vienna Ab-initio Simulation Package (VASP)~\cite{Kresse1996CMS, Kresse1996PRB}.
A plane-wave basis with a cutoff energy of 500 eV is applied to expand Kohn-Sham states.
The generalized-gradient approximation (GGA) parametrized by Perdew, Burke, and Ernzerhof (PBE) is used for the exchange-correlation functional~\cite{Perdew1996}.
All simulations are carried out with spin-polarized DFT which does not include the effect of spin-orbit coupling (SOC).
We use different $\mathbf{k}$-point grids to sample the Brillouin zone (BZ) of the different simulation cells:
These parameters are determined based on the convergence of the BCC Fe unit-cell for which a $12 \times 12 \times 12$ MP $\mathbf{k}$-grid is enough to achieve convergence of the total energy per atom to within 2 meV.
For larger supercells, we maintain a similar $\mathbf{k}$-grid density by scaling the sampling points inversely with the supercell size.
For the supercell of ideal BCC Fe and its thermally disordered lattices, we use a $8 \times 8 \times 8$ Monkhorst-Pack (MP) grid~\cite{Monkhorst1976}.
For the B doped crystalline Fe structures, Fe$_{n_\mathrm{Fe}}$B$_{n_\mathrm{B}}$, a $6 \times 6 \times 6$ MP grid is applied, and we tested that this converges the total energy to within 1 meV/atom.
For amorphous structures, we use a $3 \times 3 \times 3$ MP grid.
Due to the reduced symmetry of a-Fe, a $4 \times 4 \times 4$ MP $\mathbf{k}$-grid is computationally too expensive and we use $3 \times 3 \times 3$ points.

Structural relaxations for ideal BCC Fe and B doped BCC Fe are performed with a force tolerance of 5 meV/\AA.
We first relax undoped a-Fe structures within classical MD to a force tolerance of 0.01 eV/\AA \ and refer to these as ``MD relaxed" a-Fe.
They are then further relaxed by DFT, labeled as ``MD+DFT relaxed" a-Fe, with a force tolerance of 0.15 eV/\AA, which is the level that we can achieve within a reasonable computational time.
We generate B doped a-Fe from the ``MD relaxed'' geometries and relax these within DFT to the same force tolerance of 0.15 eV/\AA.
Thermally disordered lattices are not relaxed to maintain the frozen snapshots of the vibrating lattice.

\section{\label{sec3:results}Results and discussion}

\subsection{\label{sec3-1:phonon-results}Thermally induced atomic disorder}
\begin{figure}
\includegraphics[width=0.95\columnwidth]{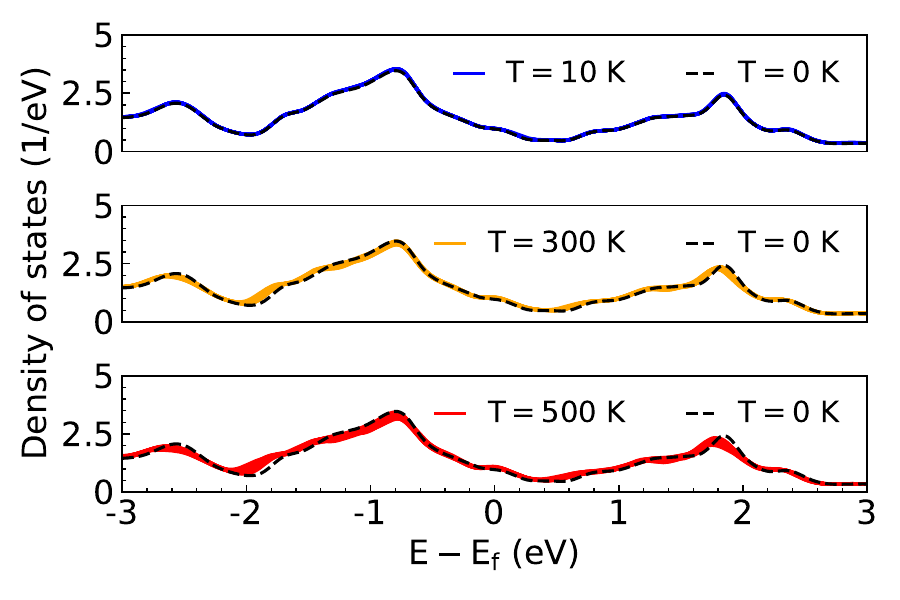}
\caption{\label{fig-2:phonon_dos}
Electronic density of states (EDOS) of pure Fe, normalized per atom, at different lattice temperatures of $T=10$~K, 300~K, and 500~K.
For each temperature, the EDOS of 50 snapshots is plotted.
The EDOS of perfect BCC Fe at $T=0$~K is included in every subplot for comparison.
}
\end{figure}

First, we investigate the effect of finite lattice temperature and the resulting atomic disorder on the EDOS of Fe for temperatures of $T=10$~K, 300~K, and 500~K.
Phonon excitations displace nuclei from their equilibrium positions, which increasingly disorders atomic geometries with increasing temperature.
The EDOS results for all 50 calculated snapshots are plotted in Fig.~\ref{fig-2:phonon_dos} for each temperature and compared to the EDOS of ideal BCC Fe at $T=0$~K. 

These results show that increasing the temperature generally broadens the linewidth of the EDOS relative to the zero-temperature counterpart due to the randomness in the atomic positions for the disordered lattice structure.
For $T=300$ and 500~K, noticeable changes to the EDOS are observed especially near peaks at $-2.6$ eV, $-0.8$ eV, and 1.8 eV as well as valleys near $-2.0$ eV and 0.6 eV, see Fig.\ \ref{fig-2:phonon_dos}.
While the density of states at the Fermi level $n(E_f)$ (EDOS-FL) is much less affected,
we provide a quantitative analysis of these changes since they may directly influence electron-magnon damping.
Towards this, we compute the average and standard deviation of EDOS-FL for each temperature and find for $T=10$~K that $n(E_f)=0.991 \pm 0.004$ eV$^{-1}$, for $T=300$~K $n(E_f)=1.038 \pm 0.012$ eV$^{-1}$, and for $T=500$~K $n(E_f)=1.058 \pm 0.019$ eV$^{-1}$.
These three values are all larger than $n(E_f)=0.974$~eV$^{-1}$ for ideal BCC Fe at zero temperature, showing that increasing temperature increases EDOS-FL for Fe.
However, even the increase of about $8.6\,\%$ at $T=500$~K is not significant, compared to the influence of doping and amorphization that we discuss later.

While our results show that EDOS-FL of Fe does not significantly change at low temperatures, this does not imply that magnetic damping is temperature independent, since mechanisms other than conductivity like damping also contribute.
Conductivity-like electron-magnon damping in the intra-band scattering limit can be approximately described by an empirical formula $\alpha_\text{intra} \sim n(E_f) |\Gamma^{-}| \tau$~\cite{Li2019AnisoDP, Fahnle2008, Gilmore2007, Kambersky2007}, where $|\Gamma^{-}|$ is the strength of the spin–orbit interaction near the Fermi level and $\tau$ is the electron relaxation time from the Drude model.
The proportionality of $\alpha_\text{intra}$ to the electron relaxation time $\tau$ implies that conductivity-like damping~\cite{Gilmore2007, Kambersky2007} is more pronounced at lower temperatures or in cleaner crystals where $\tau$ is large. 
This somewhat counter-intuitive increase of magnetic damping with increasing $\tau$ has been experimentally observed in a recent work on Fe~\cite{Khodadadi2020} and an earlier work on cobalt (Co) and Nickel (Ni)~\cite{Bhagat1974}. 
We note that in this work, we only study the dependence of $\alpha_\text{intra}$ on $n(E_f)$, however, the relaxation time also can dependent on the phonon temperature or lattice disorder, including defects.
Our simulation results clearly indicate that the EDOS-FL, one of the factors determining the intrinsic magnetic damping in metallic magnets, is not significantly affected by low temperatures.
In addition to conductivity like damping, resistivity like damping may also be affected by temperature, but is not discussed in this study.

\subsection{\label{sec3-2:doping-results}Boron doping}
\begin{figure}
\includegraphics[width=0.95\columnwidth]{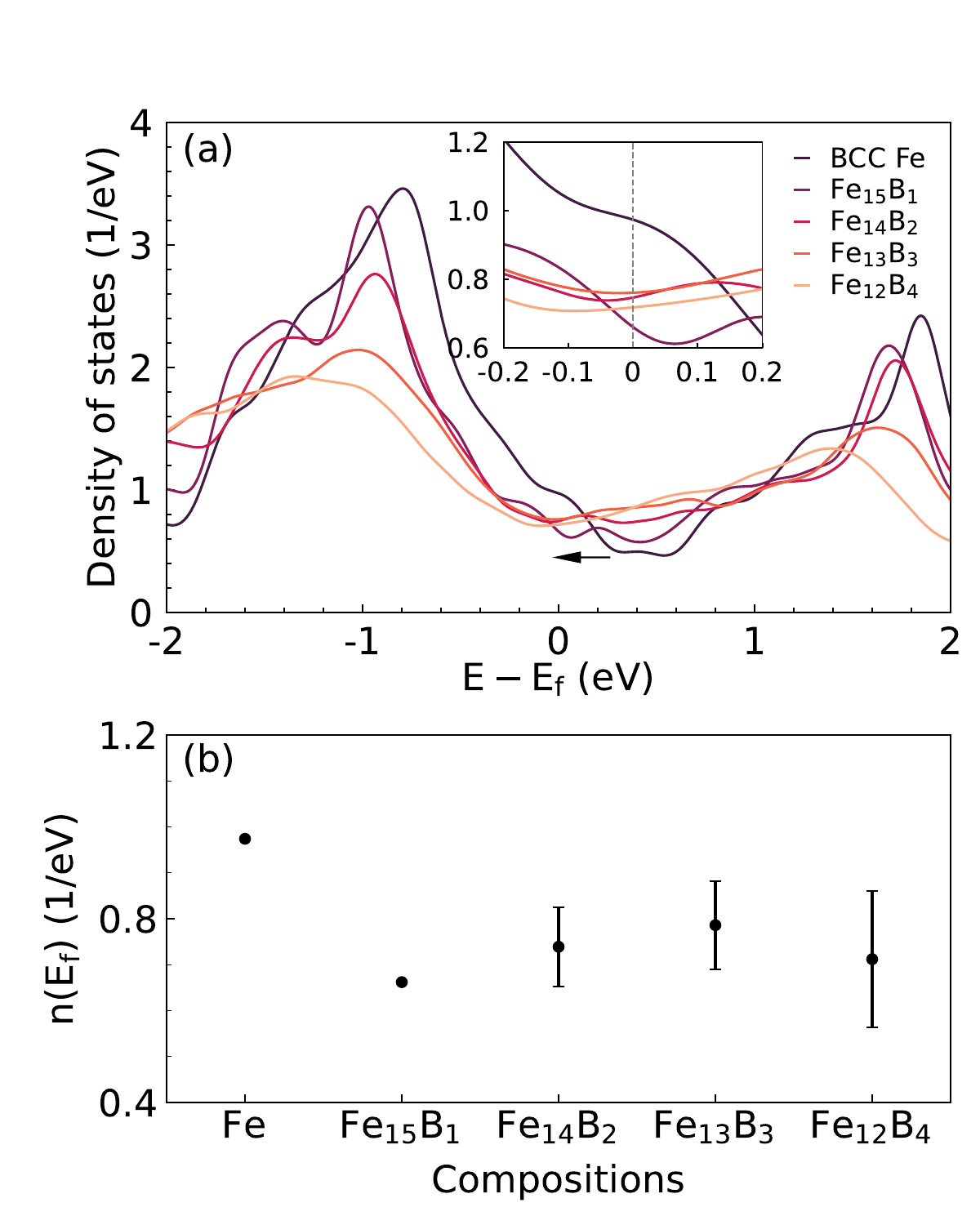}
\caption{\label{fig-3:doping_dos}
(a) Electronic density of states (EDOS), normalized per atom, for Fe and Fe doped with B, simulated using a $2 \times 2 \times 2$ supercell with different numbers of B atoms, Fe$_{16-n_\mathrm{B}}$B$_{n_\mathrm{B}}$, for $n_\mathrm{B}$=1, 2, 3, and 4.
Cluster expansion averages are shown for $n_\mathrm{B}>$1.
The inset magnifies EDOS-FL and clearly shows a reduction by B doping.
(b) Average and error bar of EDOS-FL for different doping concentrations Fe$_{16-n_\mathrm{B}}$B$_{n_\mathrm{B}}$. 
For pure Fe and $\mathrm{Fe_{15}B_1}$ only one non-equivalent class (see Table.~\ref{table1:cluster expansion}) is used and, thus, we compute no standard deviation.
}
\end{figure}

Next, we compare the EDOS of BCC Fe doped with B to that of pure BCC Fe.
Four different doping levels are considered and modeled by the structures $\mathrm{Fe_{15}B_1}$ ($6.25\%$ B), $\mathrm{Fe_{14}B_2}$ ($12.5\%$ B), $\mathrm{Fe_{13}B_3}$ ($18.75\%$ B), and $\mathrm{Fe_{12}B_4}$ ($25\%$ B), constructed using the cluster expansion method (see Sec.\ \ref{sec2:methods}).
Except for the case $n_\mathrm{B}=1$ which is represented by only one class, we calculate weighted average and standard deviation (error bar) over all classes in Table \ref{table1:cluster expansion} for the other doping levels. 

We find that doping B into BCC Fe leads to a reduced EDOS-FL compared to pure BCC Fe for all doping levels considered here [see Fig.\ \ref{fig-3:doping_dos}(a)].
This is because doping with B shifts the entire EDOS to lower energies, such that the Fermi level shifts to the dip in the original non-doped EDOS, as indicated by the black arrow in Fig.~\ref{fig-3:doping_dos}(a).
Main peaks in the EDOS, i.e.\ at $-0.784$ eV and $1.853$ eV, are also shifted to lower energies with increasing B doping. 
Meanwhile, peak intensities are decreased and peaks widths are broadened with increasing B doping.

\begin{figure}
\includegraphics[width=0.95\columnwidth]{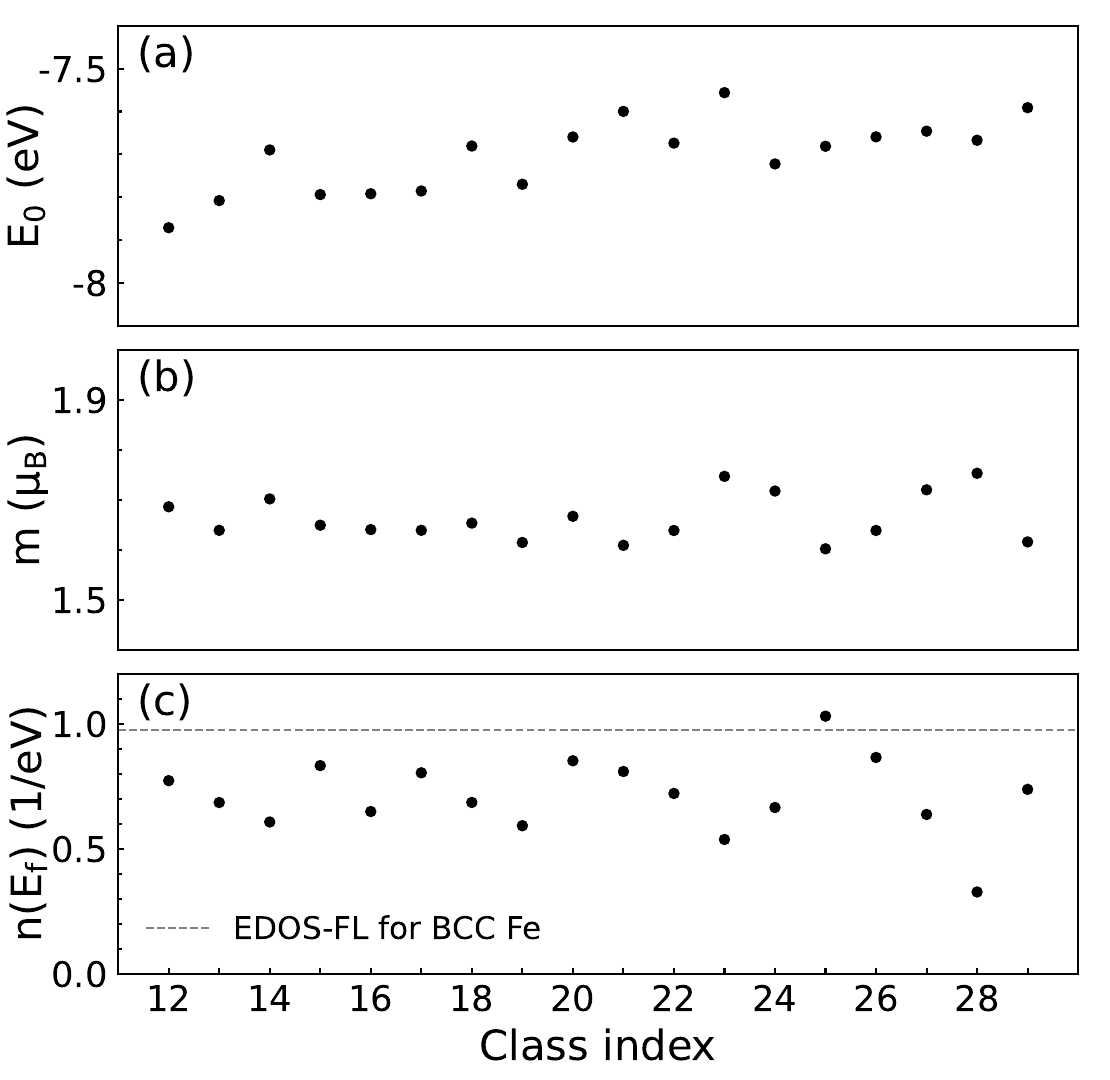}
\caption{\label{fig-4:Fe12B4_details}
(a) Ground-state total energy $E_0$ in eV, (b) magnetic moment $m$ in $\mu_\mathrm{B}$, and (c) electronic density of states at the Fermi level $n(E_f)$, normalized per atom, for all 18 non-equivalent classes corresponding to the structure Fe$_{12}$B$_{4}$ (see Table \ref{table1:cluster expansion}).
The horizontal dashed line in (c) indicates $n(E_f)$ for pure BCC Fe.
}
\end{figure}

Our data also points to a non-monotonic dependence of $n(E_f)$ on the B doping concentration and, as a result, the possibility of optimizing the doping level to minimize magnon damping.
Figure \ref{fig-3:doping_dos}(b) shows the average value and the error bar for different doped systems and pure BCC Fe.
Our results indicate an optimal B doping level of $6.25\%$, which leads to the lowest $n(E_f)=0.662$~eV$^{-1}$.
At the same time, with increasing number of B atoms in the doped cell, there are more non-equivalent classes as seen from Table~\ref{table1:cluster expansion}, and the error bar increases from $n_\mathrm{B}$=2 to $n_\mathrm{B}$=4.
In particular, EDOS-FL for Fe$_{12}$B$_4$ has a large error bar, which implies that the value of $n(E_f)$ for the 18 non-equivalent classes (see Table \ref{table1:cluster expansion}) spreads over a large range.
These classes, although having the same composition, are microscopically quite different regarding the distribution of B on the Fe lattice.
In Fig.~\ref{fig-4:Fe12B4_details}, we analyze the ground state total energy $E_0$, the magnetic moment $\vec{m}$, and EDOS-FL $n(E_f)$ for the 18 non-equivalent classes.
While $n(E_f)$ spans over a big range from 0.327 eV$^{-1}$ to 1.030 eV$^{-1}$, all values remain smaller than the value of the pure BCC Fe, except for the class with the index 25 (see Table \ref{table1:cluster expansion}).
This confirms that doping B still decreases EDOS-FL in general. 
The difference between maximum and minimum value of $E_0$, $\vec{m}$, and $n(E_f)$ amounts to 4.09\,\%, 9.08\,\%, and 98.74\,\%, respectively, relative to their average values.
We do not identify obvious correlation between the total energy and EDOS-FL in Fig.\ \ref{fig-4:Fe12B4_details}, with a correlation coefficient of $-0.110$.
For the lowest-energy class with index of $n=12$, $n(E_f)=0.772$~eV$^{-1}$, which is larger than the average, while the class with index 28 has a relatively large energy, but the lowest EDOS-FL. 
However, we do find noticeable negative correlation between the magnetic moment and EDOS-FL with a correlation coefficient of $-0.683$.
This observation is benefitial for real magnetic devices, where lower magnetic damping and higher magnetization is often desirable for applications.
It could also suggest that growing B doped Fe under an external magnetic field, and thus making high magnetic moment atomic geometries more likely, could lead to samples with lower magnetic damping.  It is already well established that synthesis or annealing of amorphous ferromagnets in applied magnetic fields may lead to induced magnetic anisotropies due to modifications of the short-range order \cite{Berry1975}.  Similar, it is conceivable that applying fields during growth or during a post-annealing may stabilize larger saturation magnetizations \cite{Wu2016}.

\subsection{\label{sec3-3:amorphous-Fe-results}Structure amorphization}
\begin{figure}
\includegraphics[width=0.95\columnwidth]{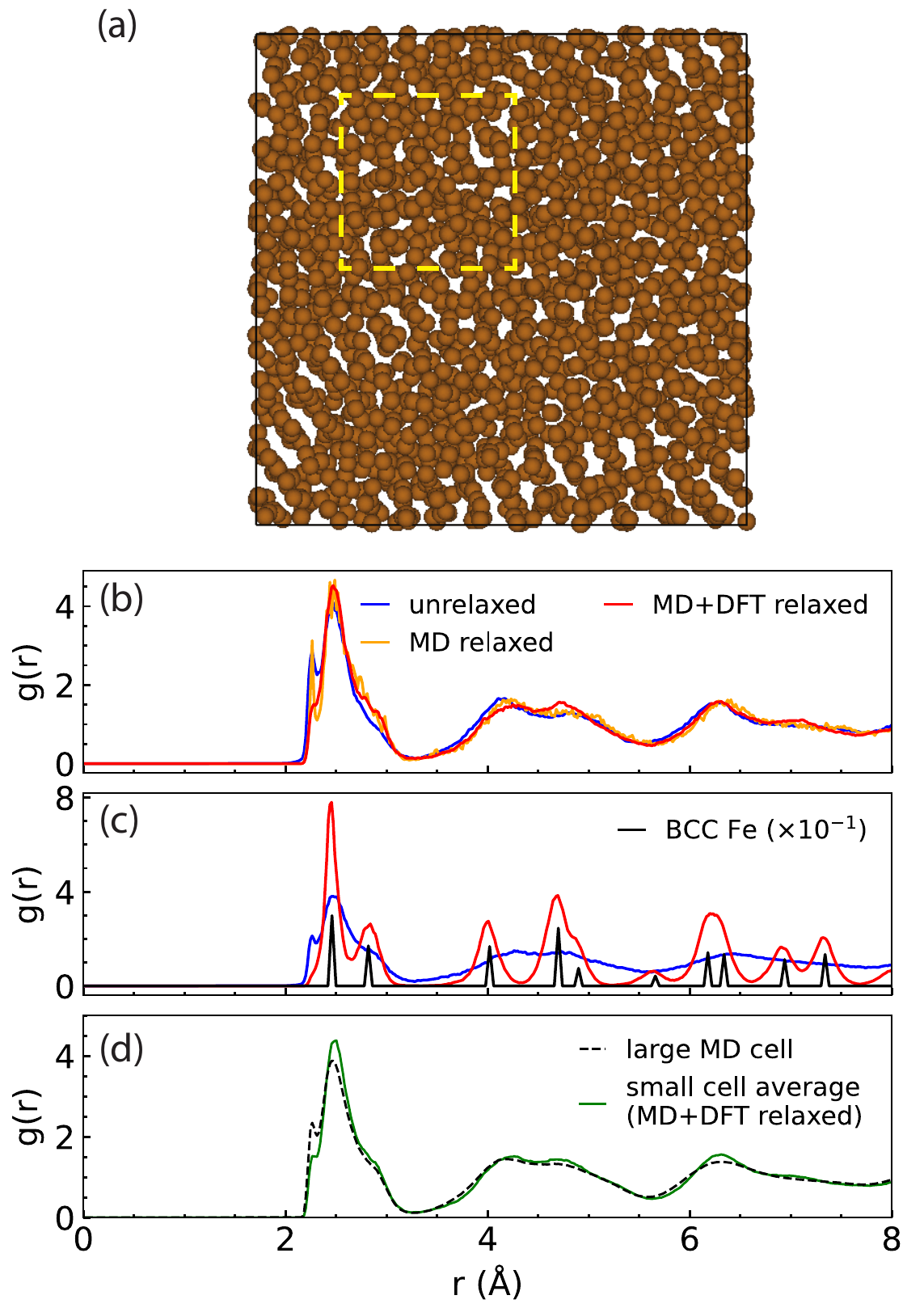}
\caption{\label{fig-5:amorphous_gr}
Amorphous Fe (a-Fe) structure and the radial distribution function (RDF) $g(r)$.
(a) Illustration of the large cubic a-Fe supercell constructed by MD.
The yellow dashed box shows a small cubic a-Fe cell cut out for DFT simulations.
(b) One representative small a-Fe cell for which DFT relaxation does not change the RDF away from the amorphous state.
(c) One representative small a-Fe cell for which DFT relaxation leads to re-crystallization, i.e., the RDF approaches the RDF of ideal BCC Fe.
(d) RDF for the large a-Fe cell created from MD, compared to the average g(r) of eight small a-Fe cells that are randomly cut from it.
}
\end{figure}

Next, we study the influence of structure amorphization by characterizing a-Fe via the radial distribution function (RDF) $g(r)$.
As explained in Sec.\ \ref{sec2:methods}, we use molecular dynamics (MD) to compute atomic positions for a large a-Fe supercell, from which we randomly cut ten small cubic cells [see yellow box in Fig.\ \ref{fig-5:amorphous_gr}(a)] and subsequently relax using MD (``MD-relaxed") or MD and DFT (``MD+DFT relaxed").
For eight of these cells, MD and DFT relaxation does not change the RDF and maintains amorphousness, while the other two relax and recrystallize, as evidenced by their RDF.
This is shown in Fig.\ \ref{fig-5:amorphous_gr}(b) for one representative a-Fe cell for which relaxation does not change the amorphousness but simply smoothens $g(r)$.
In contrast, Fig.\ \ref{fig-5:amorphous_gr}(c) shows one example for which DFT relaxation recrystallizes the amorphous structure and discrete sharp peaks appear that coincide with the RDF of ideal BCC Fe.
For our purpose of studying EDOS-FL for a-Fe, we exclude the two recrystallized samples.
The averaged RDF of the remaining eight a-Fe cells is calculated and compared in Fig.\ \ref{fig-5:amorphous_gr}(d) to the RDF of the large initial a-Fe supercell, depicted in Fig.\ \ref{fig-5:amorphous_gr}(a).
We can see that for large $r$ they almost overlap.
For small $r$ around $2.5 \text{ \AA}$, the two curves slightly differ, but still show similar peak structure.
Based on this, we conclude that the average of the small a-Fe cells approximates the large a-Fe cell well enough to study EDOS-FL for a-Fe.

\begin{figure}
\includegraphics[width=0.95\columnwidth]{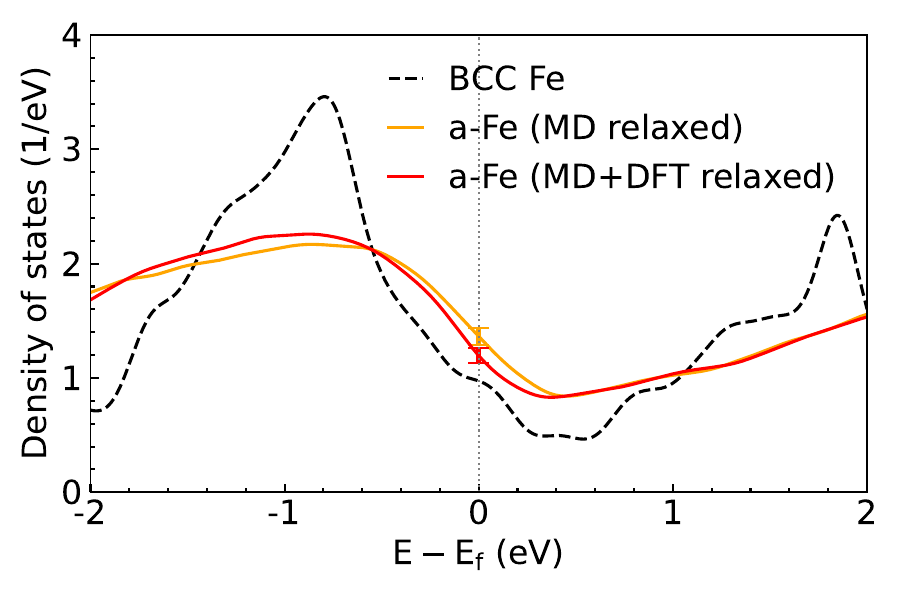}
\caption{\label{fig-6:amorphous_dos}
Electronic density of states (EDOS), normalized per atom, averaged over 8 simulation cells for amorphous Fe (a-Fe).
EDOS-FL increases for both the MD relaxed structures and ``MD+DFT" relaxed structures, relative to ideal Fe.
The error bar is computed from the standard deviation.
}
\end{figure}

Figure \ref{fig-6:amorphous_dos} shows that a-Fe has a broadened distribution of the average EDOS and an increased value of EDOS-FL, compared to ideal BCC Fe.
The peak structure of the EDOS of BCC Fe is lost for a-Fe and the entire EDOS  is smoother and broadened.
We note that performing a DFT relaxation on top of the MD relaxed structure can decrease $n(E_f)$ even further, see Fig.\ \ref{fig-6:amorphous_dos}.
However, the increase of EDOS-FL for a-Fe is larger than the error bar for both structure relaxation approaches.
Since DFT relaxations with tight force convergence criteria are computationally expensive, we do not explore this in more detail in this work.
Instead, we conclude that it is not desirable to reduce $n(E_f)$ of Fe by structure amorphization.

Finally, our $g(r)$ results illustrate that the a-Fe structures generated from our MD simulations are fundamentally different from the Fe structure up to a temperature of $T$=500 K, even though in both cases the original perfect BCC Fe lattice structure is appreciably disordered (see details in Fig.\ S2 of SI). 
Comparing to the $g(r)$ of the perfect BCC Fe at zero lattice temperature, the temperature induced disorder just broadens the original peaks.
Some individual peaks are merged if they are too close to each other along the $r$-axis.
The a-Fe structure, however, has a fundamentally different $g(r)$.
A new distinct peak appears below the lowest inter-atomic distance of BCC Fe, which is even lying outside the broadening observed for $T$=500 K.
Meanwhile, $g(r)$ for larger $r$ has a nearly continuous distribution instead of a discrete distribution sharply peaked at specific distances.
This is consistent with our result of only a minor influence of lattice temperature on EDOS-FL and a more noticeable change for a-Fe.

\subsection{\label{sec3-4:amorphous-FeB-results}Amorphous iron with boron doping}
\begin{figure}
\includegraphics[width=0.95\columnwidth]{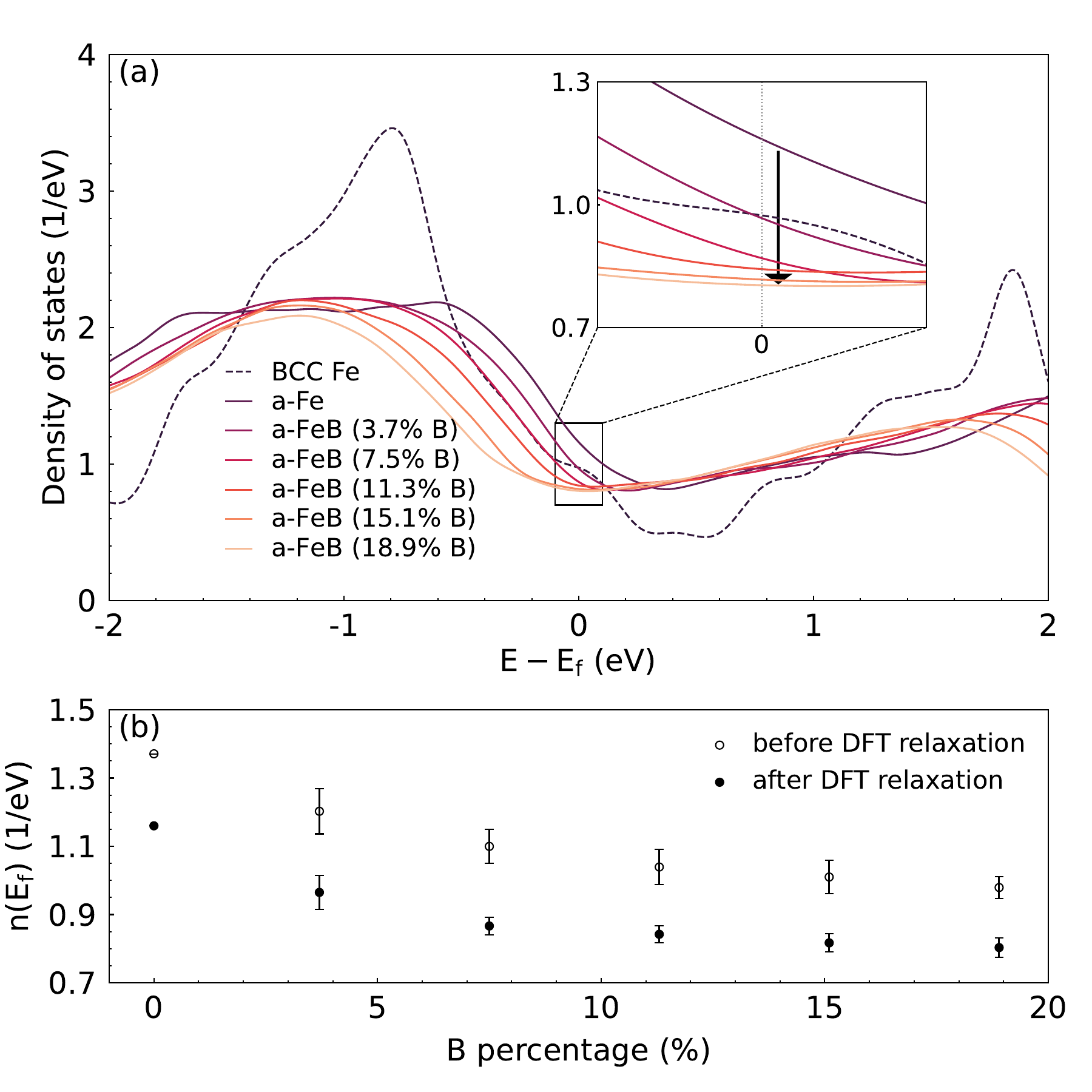}
\caption{\label{fig-7:amorphous_FeB_dos}
(a) EDOS per atom for ideal BCC Fe (same data as in Fig.\ \ref{fig-3:doping_dos}), a-Fe (the single snapshot used to build doped a-Fe through random replacement), and a-FeB (B doped a-Fe) with different doping levels of 3.7\,\%, 7.5\,\%, 11.3\,\%, 15.1\,\%, and 18.9\,\%.
For each doping level of a-FeB the plotted result is the average over 10 different snapshots with random replacements of Fe atoms by B atoms, starting from a single original a-Fe supercell.
The inset magnifies the EDOS near the Fermi level.
(b) Average and standard deviation of EDOS-FL for different levels of B doping.
Results before DFT relaxation are shown for comparison. 
}
\end{figure}

Finally, after showing in Sec.~\ref{sec3-2:doping-results} that B doping can decrease the electronic density of states at the Fermi level (EDOS-FL) for \textit{crystalline} Fe, we now consider B doping of a-Fe. 
For each doping level, the random distribution of B sites is considered by calculating ten possible configurations.
They are constructed as described in Sec.\ \ref{sec2:methods} and subsequently relaxed using DFT until forces on all atoms are smaller than $0.15\ \text{eV/\AA}$, i.e., the same force tolerance used for ``MD+DFT relaxed" a-Fe.
For each of these the EDOS is computed and averaged for analysis.

Our results in Fig.~\ref{fig-7:amorphous_FeB_dos}(a) show that B doping decreases the averaged EDOS at the Fermi level $n(E_f)$ (EDOS-FL) for a-FeB, similar to what we observed for crystalline BCC Fe.
With increasing B doping, the EDOS-FL of a-FeB monotonically decreases in the range from 0\,\% to 18.9\,\% B, and starts to saturate around 15\,\%.
The inset in Fig.\ \ref{fig-7:amorphous_FeB_dos}(a) focuses on this trend near the Fermi energy (see black arrow).
Interestingly, doping B into a-Fe can lead to an EDOS-FL that is even smaller than that of pure BCC Fe, overcoming the increase we discussed for a-Fe (see Fig.\ \ref{fig-6:amorphous_dos}).
This implies that a-FeB is potentially advantageous for reducing intrinsic intra-band magnon-electron damping, even though undoped a-Fe is not.

This is illustrated more clearly in Fig.\ \ref{fig-7:amorphous_FeB_dos}(b) via average and standard deviation.
The decrease of $n(E_f)$ with increasing B doping from 0 to 18.9\,\% is outside the statistical error bars and gradually saturates when the doping percentage is $\gtrsim 15\,\%$. 
Therefore, we do not consider larger doping percentages in this work.
It is possible that further increasing the doping percentage may even increase $n(E_f)$ similar to what we have observed for B doped BCC Fe in Fig.\ \ref{fig-3:doping_dos}(b).
Exploring this is outside the scope of this work, since it is
undesired for reducing magnetic damping. 
Our simulations indicate that $18.9\,\%$ of B doping is near the optimal doping level that corresponds to the lowest $n(E_f)$ for a-FeB.

Finally, for all simulated a-FeB structures, DFT relaxation leads to a reduction of EDOS-FL, similar to what we observed for a-Fe in Fig.\ \ref{fig-6:amorphous_dos}. 
Figure \ref{fig-7:amorphous_FeB_dos}(b) shows for different B doping percentages that this reduction is beyond the range of the statistical error bar.
The trend in this figure indicates that while relaxing the amorphous structure to even lower force tolerance is computationally costly, it could lead to slightly lower values of EDOS-FL.

\begin{table}
\begin{tabular}{|c|c|c|c|} 
\hline
\multicolumn{2}{|c|}{Structures} & EDOS-FL ($\mathrm{1/eV}$) & Error \\
\hline
\multirow{4}{*}{BCC Fe} & $T$=0 K & 0.974 & --- \\ \cline{2-4}
                        & $T$=10 K & 0.991 & 0.004 \\ \cline{2-4} 
                        & $T$=300 K & 1.038 & 0.012 \\ \cline{2-4}
                        & $T$=500 K & 1.058 & 0.019 \\ \hline
\multirow{4}{*}{B-doped BCC Fe} & 6.25\,\% B & 0.662 & --- \\ \cline{2-4}
                        & 12.50\,\% B & 0.739 & 0.086 \\ \cline{2-4} 
                        & 18.75\,\% B & 0.786 & 0.096 \\ \cline{2-4}
                        & 25.00\,\% B & 0.712 & 0.149 \\ \hline
\multicolumn{2}{|c|}{a-Fe} & 1.199 & 0.065 \\ \hline
\multirow{5}{*}{a-FeB} & 3.7\,\% B & 0.965 & 0.049 \\ \cline{2-4}
                        & 7.5\,\% B & 0.866 & 0.025 \\ \cline{2-4} 
                        & 11.3\,\% B & 0.842 & 0.024 \\ \cline{2-4}
                        & 15.1\,\% B & 0.817 & 0.027 \\ \cline{2-4}
                        & 18.9\,\% B & 0.803 & 0.029 \\ \hline
\end{tabular}
\caption{\label{table2:edos_summary}
Electron density of states of BCC Fe at the Fermi level (EDOS-FL) for all different structures studied in this work, i.e., non-zero lattice temperature, B doping, amorphous lattice (a-Fe), and B doped amorphous Fe (a-FeB).
Values are normalized per atom and we report averages and standard deviations for the cases where multiple configurations contribute (see text).
}
\end{table}

We summarize our numerical results for averages and standard deviations of EDOS-FL for all simulated structures in Table~\ref{table2:edos_summary}.
This shows that crystalline BCC Fe doped with 6.25\,\% B leads to the lowest EDOS-FL among all structures considered.
Previous experimental work demonstrated that doping carbon into Co-Fe alloys can naturally make the structure amorphous and that amorphous Co-Fe-C alloys can have smaller magnetic damping than polycrystalline Co-Fe alloys~\cite{Wang2019}.
Recent experiments confirmed that similar effects can be achieved with B doping \cite{Lourembam2021}.
While this is explained by smaller grain sizes in the amorphous material and, therefore, reduced sample inhomogeneity leading to lower damping, our results indicate that B doping of amorphous Fe is an additional factor that can reduce the intrinsic magnetic damping by decreasing the EDOS-FL.

\section{\label{sec4:conclusion}Conclusions}
We used first-principles simulations to quantitatively study the influence of lattice temperature, boron doping, and structure amorphization on the electronic structure of iron.
Using our results, we discussed how these factors affect the electronic density of states at the Fermi level, as one important quantity determining the intrinsic magnetic damping in metallic magnets within Kambersk{\'y}'s breathing Fermi surface model.
Generally, we showed that both structural effects modify the density of states near the Fermi level to a lesser extent.
Instead, doping with B is advantageous in crystalline and amorphous Fe to reduce damping and an optimal doping level of 6.25\,\% B reduces EDOS-FL by 32\,\% in crystalline Fe. 
We also showed that different arrangements of the B atoms on Fe sites affect the resulting magnetization and EDOS-FL values differently.
This possibly suggests materials growth under an external magnetic field to favor arrangements with large magnetization and small damping.
While this work points to a reduction of EDOS-FL as one reason for reduced magnetic damping in B doped Fe, we expect similar mechanisms to govern C doped material and Co-Fe alloys.

\begin{acknowledgments}
We thank Brian R.\ Robinson for insightful discussions and valuable suggestion during the development of this manuscript.
This work was supported by the U.S.\ DOE, Office of Science, Basic Energy Sciences, Materials Sciences and Engineering Division under contract No.\ DE-SC0022060.
This work made use of the Illinois Campus Cluster, a computing resource that is operated by the Illinois Campus Cluster Program (ICCP) in conjunction with the National Center for Supercomputing Applications (NCSA) and which is supported by funds from the University of Illinois at Urbana-Champaign.
\end{acknowledgments}


\bibliography{refs}

\begin{thebibliography}{60}%
\makeatletter
\providecommand \@ifxundefined [1]{%
 \@ifx{#1\undefined}
}%
\providecommand \@ifnum [1]{%
 \ifnum #1\expandafter \@firstoftwo
 \else \expandafter \@secondoftwo
 \fi
}%
\providecommand \@ifx [1]{%
 \ifx #1\expandafter \@firstoftwo
 \else \expandafter \@secondoftwo
 \fi
}%
\providecommand \natexlab [1]{#1}%
\providecommand \enquote  [1]{``#1''}%
\providecommand \bibnamefont  [1]{#1}%
\providecommand \bibfnamefont [1]{#1}%
\providecommand \citenamefont [1]{#1}%
\providecommand \href@noop [0]{\@secondoftwo}%
\providecommand \href [0]{\begingroup \@sanitize@url \@href}%
\providecommand \@href[1]{\@@startlink{#1}\@@href}%
\providecommand \@@href[1]{\endgroup#1\@@endlink}%
\providecommand \@sanitize@url [0]{\catcode `\\12\catcode `\$12\catcode
  `\&12\catcode `\#12\catcode `\^12\catcode `\_12\catcode `\%12\relax}%
\providecommand \@@startlink[1]{}%
\providecommand \@@endlink[0]{}%
\providecommand \url  [0]{\begingroup\@sanitize@url \@url }%
\providecommand \@url [1]{\endgroup\@href {#1}{\urlprefix }}%
\providecommand \urlprefix  [0]{URL }%
\providecommand \Eprint [0]{\href }%
\providecommand \doibase [0]{https://doi.org/}%
\providecommand \selectlanguage [0]{\@gobble}%
\providecommand \bibinfo  [0]{\@secondoftwo}%
\providecommand \bibfield  [0]{\@secondoftwo}%
\providecommand \translation [1]{[#1]}%
\providecommand \BibitemOpen [0]{}%
\providecommand \bibitemStop [0]{}%
\providecommand \bibitemNoStop [0]{.\EOS\space}%
\providecommand \EOS [0]{\spacefactor3000\relax}%
\providecommand \BibitemShut  [1]{\csname bibitem#1\endcsname}%
\let\auto@bib@innerbib\@empty
\bibitem [{\citenamefont {Li}\ \emph {et~al.}(2020)\citenamefont {Li},
  \citenamefont {Zhang}, \citenamefont {Tyberkevych}, \citenamefont {Kwok},
  \citenamefont {Hoffmann},\ and\ \citenamefont {Novosad}}]{Li2020}%
  \BibitemOpen
  \bibfield  {author} {\bibinfo {author} {\bibfnamefont {Y.}~\bibnamefont
  {Li}}, \bibinfo {author} {\bibfnamefont {W.}~\bibnamefont {Zhang}}, \bibinfo
  {author} {\bibfnamefont {V.}~\bibnamefont {Tyberkevych}}, \bibinfo {author}
  {\bibfnamefont {W.~K.}\ \bibnamefont {Kwok}}, \bibinfo {author}
  {\bibfnamefont {A.}~\bibnamefont {Hoffmann}},\ and\ \bibinfo {author}
  {\bibfnamefont {V.}~\bibnamefont {Novosad}},\ }\bibfield  {title} {\bibinfo
  {title} {Hybrid magnonics: Physics, circuits, and applications for coherent
  information processing},\ }\href {https://doi.org/10.1063/5.0020277/287124}
  {\bibfield  {journal} {\bibinfo  {journal} {Journal of Applied Physics}\
  }\textbf {\bibinfo {volume} {128}},\ \bibinfo {pages} {130902} (\bibinfo
  {year} {2020})}\BibitemShut {NoStop}%
\bibitem [{\citenamefont {Awschalom}\ \emph {et~al.}(2021)\citenamefont
  {Awschalom}, \citenamefont {Du}, \citenamefont {He}, \citenamefont
  {Heremans}, \citenamefont {Hoffmann}, \citenamefont {Hou}, \citenamefont
  {Kurebayashi}, \citenamefont {Li}, \citenamefont {Liu}, \citenamefont
  {Novosad}, \citenamefont {Sklenar}, \citenamefont {Sullivan}, \citenamefont
  {Sun}, \citenamefont {Tang}, \citenamefont {Tyberkevych}, \citenamefont
  {Trevillian}, \citenamefont {Tsen}, \citenamefont {Weiss}, \citenamefont
  {Zhang}, \citenamefont {Zhang}, \citenamefont {Zhao},\ and\ \citenamefont
  {Zollitsch}}]{Awschalom2021}%
  \BibitemOpen
  \bibfield  {author} {\bibinfo {author} {\bibfnamefont {D.~D.}\ \bibnamefont
  {Awschalom}}, \bibinfo {author} {\bibfnamefont {C.~R.}\ \bibnamefont {Du}},
  \bibinfo {author} {\bibfnamefont {R.}~\bibnamefont {He}}, \bibinfo {author}
  {\bibfnamefont {F.~J.}\ \bibnamefont {Heremans}}, \bibinfo {author}
  {\bibfnamefont {A.}~\bibnamefont {Hoffmann}}, \bibinfo {author}
  {\bibfnamefont {J.}~\bibnamefont {Hou}}, \bibinfo {author} {\bibfnamefont
  {H.}~\bibnamefont {Kurebayashi}}, \bibinfo {author} {\bibfnamefont
  {Y.}~\bibnamefont {Li}}, \bibinfo {author} {\bibfnamefont {L.}~\bibnamefont
  {Liu}}, \bibinfo {author} {\bibfnamefont {V.}~\bibnamefont {Novosad}},
  \bibinfo {author} {\bibfnamefont {J.}~\bibnamefont {Sklenar}}, \bibinfo
  {author} {\bibfnamefont {S.~E.}\ \bibnamefont {Sullivan}}, \bibinfo {author}
  {\bibfnamefont {D.}~\bibnamefont {Sun}}, \bibinfo {author} {\bibfnamefont
  {H.}~\bibnamefont {Tang}}, \bibinfo {author} {\bibfnamefont {V.}~\bibnamefont
  {Tyberkevych}}, \bibinfo {author} {\bibfnamefont {C.}~\bibnamefont
  {Trevillian}}, \bibinfo {author} {\bibfnamefont {A.~W.}\ \bibnamefont
  {Tsen}}, \bibinfo {author} {\bibfnamefont {L.~R.}\ \bibnamefont {Weiss}},
  \bibinfo {author} {\bibfnamefont {W.}~\bibnamefont {Zhang}}, \bibinfo
  {author} {\bibfnamefont {X.}~\bibnamefont {Zhang}}, \bibinfo {author}
  {\bibfnamefont {L.}~\bibnamefont {Zhao}},\ and\ \bibinfo {author}
  {\bibfnamefont {C.~H.}\ \bibnamefont {Zollitsch}},\ }\bibfield  {title}
  {\bibinfo {title} {Quantum engineering with hybrid magnonic systems and
  materials},\ }\bibfield  {journal} {\bibinfo  {journal} {IEEE Transactions on
  Quantum Engineering}\ }\textbf {\bibinfo {volume} {2}},\ \href
  {https://doi.org/10.1109/TQE.2021.3057799} {10.1109/TQE.2021.3057799}
  (\bibinfo {year} {2021})\BibitemShut {NoStop}%
\bibitem [{\citenamefont {Yuan}\ \emph {et~al.}(2022)\citenamefont {Yuan},
  \citenamefont {Cao}, \citenamefont {Kamra}, \citenamefont {Duine},\ and\
  \citenamefont {Yan}}]{Yuan2022}%
  \BibitemOpen
  \bibfield  {author} {\bibinfo {author} {\bibfnamefont {H.~Y.}\ \bibnamefont
  {Yuan}}, \bibinfo {author} {\bibfnamefont {Y.}~\bibnamefont {Cao}}, \bibinfo
  {author} {\bibfnamefont {A.}~\bibnamefont {Kamra}}, \bibinfo {author}
  {\bibfnamefont {R.~A.}\ \bibnamefont {Duine}},\ and\ \bibinfo {author}
  {\bibfnamefont {P.}~\bibnamefont {Yan}},\ }\bibfield  {title} {\bibinfo
  {title} {Quantum magnonics: When magnon spintronics meets quantum information
  science},\ }\href {https://doi.org/10.1016/J.PHYSREP.2022.03.002} {\bibfield
  {journal} {\bibinfo  {journal} {Physics Reports}\ }\textbf {\bibinfo {volume}
  {965}},\ \bibinfo {pages} {1} (\bibinfo {year} {2022})}\BibitemShut {NoStop}%
\bibitem [{\citenamefont {Jiang}\ \emph {et~al.}(2023)\citenamefont {Jiang},
  \citenamefont {Lim}, \citenamefont {Li}, \citenamefont {Pfaff}, \citenamefont
  {Lo}, \citenamefont {Qian}, \citenamefont {Schleife}, \citenamefont {Zuo},
  \citenamefont {Novosad},\ and\ \citenamefont {Hoffmann}}]{Jiang2023}%
  \BibitemOpen
  \bibfield  {author} {\bibinfo {author} {\bibfnamefont {Z.}~\bibnamefont
  {Jiang}}, \bibinfo {author} {\bibfnamefont {J.}~\bibnamefont {Lim}}, \bibinfo
  {author} {\bibfnamefont {Y.}~\bibnamefont {Li}}, \bibinfo {author}
  {\bibfnamefont {W.}~\bibnamefont {Pfaff}}, \bibinfo {author} {\bibfnamefont
  {T.-H.}\ \bibnamefont {Lo}}, \bibinfo {author} {\bibfnamefont
  {J.}~\bibnamefont {Qian}}, \bibinfo {author} {\bibfnamefont {A.}~\bibnamefont
  {Schleife}}, \bibinfo {author} {\bibfnamefont {J.-M.}\ \bibnamefont {Zuo}},
  \bibinfo {author} {\bibfnamefont {V.}~\bibnamefont {Novosad}},\ and\ \bibinfo
  {author} {\bibfnamefont {A.}~\bibnamefont {Hoffmann}},\ }\bibfield  {title}
  {\bibinfo {title} {Integrating magnons for quantum information},\ }\bibfield
  {journal} {\bibinfo  {journal} {Applied Physics Letters}\ }\textbf {\bibinfo
  {volume} {123}},\ \href {https://doi.org/10.1063/5.0157520/2912767}
  {10.1063/5.0157520/2912767} (\bibinfo {year} {2023})\BibitemShut {NoStop}%
\bibitem [{\citenamefont {Tabuchi}\ \emph {et~al.}(2015)\citenamefont
  {Tabuchi}, \citenamefont {Ishino}, \citenamefont {Noguchi}, \citenamefont
  {Ishikawa}, \citenamefont {Yamazaki}, \citenamefont {Usami},\ and\
  \citenamefont {Nakamura}}]{Tabuchi2015}%
  \BibitemOpen
  \bibfield  {author} {\bibinfo {author} {\bibfnamefont {Y.}~\bibnamefont
  {Tabuchi}}, \bibinfo {author} {\bibfnamefont {S.}~\bibnamefont {Ishino}},
  \bibinfo {author} {\bibfnamefont {A.}~\bibnamefont {Noguchi}}, \bibinfo
  {author} {\bibfnamefont {T.}~\bibnamefont {Ishikawa}}, \bibinfo {author}
  {\bibfnamefont {R.}~\bibnamefont {Yamazaki}}, \bibinfo {author}
  {\bibfnamefont {K.}~\bibnamefont {Usami}},\ and\ \bibinfo {author}
  {\bibfnamefont {Y.}~\bibnamefont {Nakamura}},\ }\bibfield  {title} {\bibinfo
  {title} {Coherent coupling between a ferromagnetic magnon and a
  superconducting qubit},\ }\href
  {https://doi.org/10.1126/SCIENCE.AAA3693/SUPPL_FILE/TABUCHI-SM.PDF}
  {\bibfield  {journal} {\bibinfo  {journal} {Science}\ }\textbf {\bibinfo
  {volume} {349}},\ \bibinfo {pages} {405} (\bibinfo {year}
  {2015})}\BibitemShut {NoStop}%
\bibitem [{\citenamefont {Rezende}(2020)}]{Rezende2020}%
  \BibitemOpen
  \bibfield  {author} {\bibinfo {author} {\bibfnamefont {S.~M.}\ \bibnamefont
  {Rezende}},\ }\href {https://doi.org/10.1007/978-3-030-41317-0} {\emph
  {\bibinfo {title} {Fundamentals of magnonics}}},\ Vol.\ \bibinfo {volume}
  {969}\ (\bibinfo  {publisher} {Springer International Publishing},\ \bibinfo
  {year} {2020})\BibitemShut {NoStop}%
\bibitem [{\citenamefont {Kostylev}\ \emph {et~al.}(2016)\citenamefont
  {Kostylev}, \citenamefont {Goryachev},\ and\ \citenamefont
  {Tobar}}]{Kostylev2016}%
  \BibitemOpen
  \bibfield  {author} {\bibinfo {author} {\bibfnamefont {N.}~\bibnamefont
  {Kostylev}}, \bibinfo {author} {\bibfnamefont {M.}~\bibnamefont
  {Goryachev}},\ and\ \bibinfo {author} {\bibfnamefont {M.~E.}\ \bibnamefont
  {Tobar}},\ }\bibfield  {title} {\bibinfo {title} {Superstrong coupling of a
  microwave cavity to yttrium iron garnet magnons},\ }\bibfield  {journal}
  {\bibinfo  {journal} {Applied Physics Letters}\ }\textbf {\bibinfo {volume}
  {108}},\ \href {https://doi.org/10.1063/1.4941730/594143}
  {10.1063/1.4941730/594143} (\bibinfo {year} {2016})\BibitemShut {NoStop}%
\bibitem [{\citenamefont {Bhoi}\ \emph {et~al.}(2014)\citenamefont {Bhoi},
  \citenamefont {Cliff}, \citenamefont {Maksymov}, \citenamefont {Kostylev},
  \citenamefont {Aiyar}, \citenamefont {Venkataramani}, \citenamefont
  {Prasad},\ and\ \citenamefont {Stamps}}]{Bhoi2014}%
  \BibitemOpen
  \bibfield  {author} {\bibinfo {author} {\bibfnamefont {B.}~\bibnamefont
  {Bhoi}}, \bibinfo {author} {\bibfnamefont {T.}~\bibnamefont {Cliff}},
  \bibinfo {author} {\bibfnamefont {I.~S.}\ \bibnamefont {Maksymov}}, \bibinfo
  {author} {\bibfnamefont {M.}~\bibnamefont {Kostylev}}, \bibinfo {author}
  {\bibfnamefont {R.}~\bibnamefont {Aiyar}}, \bibinfo {author} {\bibfnamefont
  {N.}~\bibnamefont {Venkataramani}}, \bibinfo {author} {\bibfnamefont
  {S.}~\bibnamefont {Prasad}},\ and\ \bibinfo {author} {\bibfnamefont {R.~L.}\
  \bibnamefont {Stamps}},\ }\bibfield  {title} {\bibinfo {title} {Study of
  photon-magnon coupling in a yig-film split-ring resonant system},\ }\href
  {https://doi.org/10.1063/1.4904857/152080} {\bibfield  {journal} {\bibinfo
  {journal} {Journal of Applied Physics}\ }\textbf {\bibinfo {volume} {116}},\
  \bibinfo {pages} {243906} (\bibinfo {year} {2014})}\BibitemShut {NoStop}%
\bibitem [{\citenamefont {Li}\ \emph {et~al.}(2019{\natexlab{a}})\citenamefont
  {Li}, \citenamefont {Polakovic}, \citenamefont {Wang}, \citenamefont {Xu},
  \citenamefont {Lendinez}, \citenamefont {Zhang}, \citenamefont {Ding},
  \citenamefont {Khaire}, \citenamefont {Saglam}, \citenamefont {Divan},
  \citenamefont {Pearson}, \citenamefont {Kwok}, \citenamefont {Xiao},
  \citenamefont {Novosad}, \citenamefont {Hoffmann},\ and\ \citenamefont
  {Zhang}}]{Li2019PRL}%
  \BibitemOpen
  \bibfield  {author} {\bibinfo {author} {\bibfnamefont {Y.}~\bibnamefont
  {Li}}, \bibinfo {author} {\bibfnamefont {T.}~\bibnamefont {Polakovic}},
  \bibinfo {author} {\bibfnamefont {Y.~L.}\ \bibnamefont {Wang}}, \bibinfo
  {author} {\bibfnamefont {J.}~\bibnamefont {Xu}}, \bibinfo {author}
  {\bibfnamefont {S.}~\bibnamefont {Lendinez}}, \bibinfo {author}
  {\bibfnamefont {Z.}~\bibnamefont {Zhang}}, \bibinfo {author} {\bibfnamefont
  {J.}~\bibnamefont {Ding}}, \bibinfo {author} {\bibfnamefont {T.}~\bibnamefont
  {Khaire}}, \bibinfo {author} {\bibfnamefont {H.}~\bibnamefont {Saglam}},
  \bibinfo {author} {\bibfnamefont {R.}~\bibnamefont {Divan}}, \bibinfo
  {author} {\bibfnamefont {J.}~\bibnamefont {Pearson}}, \bibinfo {author}
  {\bibfnamefont {W.~K.}\ \bibnamefont {Kwok}}, \bibinfo {author}
  {\bibfnamefont {Z.}~\bibnamefont {Xiao}}, \bibinfo {author} {\bibfnamefont
  {V.}~\bibnamefont {Novosad}}, \bibinfo {author} {\bibfnamefont
  {A.}~\bibnamefont {Hoffmann}},\ and\ \bibinfo {author} {\bibfnamefont
  {W.}~\bibnamefont {Zhang}},\ }\bibfield  {title} {\bibinfo {title} {Strong
  coupling between magnons and microwave photons in on-chip
  ferromagnet-superconductor thin-film devices},\ }\href
  {https://doi.org/10.1103/PHYSREVLETT.123.107701/FIGURES/4/MEDIUM} {\bibfield
  {journal} {\bibinfo  {journal} {Physical Review Letters}\ }\textbf {\bibinfo
  {volume} {123}},\ \bibinfo {pages} {107701} (\bibinfo {year}
  {2019}{\natexlab{a}})}\BibitemShut {NoStop}%
\bibitem [{\citenamefont {Hou}\ and\ \citenamefont {Liu}(2019)}]{Hou2019}%
  \BibitemOpen
  \bibfield  {author} {\bibinfo {author} {\bibfnamefont {J.~T.}\ \bibnamefont
  {Hou}}\ and\ \bibinfo {author} {\bibfnamefont {L.}~\bibnamefont {Liu}},\
  }\bibfield  {title} {\bibinfo {title} {Strong coupling between microwave
  photons and nanomagnet magnons},\ }\href
  {https://doi.org/10.1103/PHYSREVLETT.123.107702/FIGURES/3/MEDIUM} {\bibfield
  {journal} {\bibinfo  {journal} {Physical Review Letters}\ }\textbf {\bibinfo
  {volume} {123}},\ \bibinfo {pages} {107702} (\bibinfo {year}
  {2019})}\BibitemShut {NoStop}%
\bibitem [{\citenamefont {Schmidt}\ \emph {et~al.}(2020)\citenamefont
  {Schmidt}, \citenamefont {Hauser}, \citenamefont {Trempler}, \citenamefont
  {Paleschke},\ and\ \citenamefont {Papaioannou}}]{Schmidt2020}%
  \BibitemOpen
  \bibfield  {author} {\bibinfo {author} {\bibfnamefont {G.}~\bibnamefont
  {Schmidt}}, \bibinfo {author} {\bibfnamefont {C.}~\bibnamefont {Hauser}},
  \bibinfo {author} {\bibfnamefont {P.}~\bibnamefont {Trempler}}, \bibinfo
  {author} {\bibfnamefont {M.}~\bibnamefont {Paleschke}},\ and\ \bibinfo
  {author} {\bibfnamefont {E.~T.}\ \bibnamefont {Papaioannou}},\ }\bibfield
  {title} {\bibinfo {title} {Ultra thin films of yttrium iron garnet with very
  low damping: A review},\ }\href {https://doi.org/10.1002/PSSB.201900644}
  {\bibfield  {journal} {\bibinfo  {journal} {physica status solidi (b)}\
  }\textbf {\bibinfo {volume} {257}},\ \bibinfo {pages} {1900644} (\bibinfo
  {year} {2020})}\BibitemShut {NoStop}%
\bibitem [{\citenamefont {Glass}\ and\ \citenamefont
  {Elliot}(1976)}]{Glass1976}%
  \BibitemOpen
  \bibfield  {author} {\bibinfo {author} {\bibfnamefont {H.~L.}\ \bibnamefont
  {Glass}}\ and\ \bibinfo {author} {\bibfnamefont {M.~T.}\ \bibnamefont
  {Elliot}},\ }\bibfield  {title} {\bibinfo {title} {Attainment of the
  intrinsic fmr linewidth in yttrium iron garnet films grown by liquid phase
  epitaxy},\ }\href {https://doi.org/10.1016/0022-0248(76)90141-X} {\bibfield
  {journal} {\bibinfo  {journal} {Journal of Crystal Growth}\ }\textbf
  {\bibinfo {volume} {34}},\ \bibinfo {pages} {285} (\bibinfo {year}
  {1976})}\BibitemShut {NoStop}%
\bibitem [{\citenamefont {Hirohata}\ \emph {et~al.}(2020)\citenamefont
  {Hirohata}, \citenamefont {Yamada}, \citenamefont {Nakatani}, \citenamefont
  {Prejbeanu}, \citenamefont {Diény}, \citenamefont {Pirro},\ and\
  \citenamefont {Hillebrands}}]{Hirohata2020}%
  \BibitemOpen
  \bibfield  {author} {\bibinfo {author} {\bibfnamefont {A.}~\bibnamefont
  {Hirohata}}, \bibinfo {author} {\bibfnamefont {K.}~\bibnamefont {Yamada}},
  \bibinfo {author} {\bibfnamefont {Y.}~\bibnamefont {Nakatani}}, \bibinfo
  {author} {\bibfnamefont {L.}~\bibnamefont {Prejbeanu}}, \bibinfo {author}
  {\bibfnamefont {B.}~\bibnamefont {Diény}}, \bibinfo {author} {\bibfnamefont
  {P.}~\bibnamefont {Pirro}},\ and\ \bibinfo {author} {\bibfnamefont
  {B.}~\bibnamefont {Hillebrands}},\ }\bibfield  {title} {\bibinfo {title}
  {Review on spintronics: Principles and device applications},\ }\href
  {https://doi.org/10.1016/J.JMMM.2020.166711} {\bibfield  {journal} {\bibinfo
  {journal} {Journal of Magnetism and Magnetic Materials}\ }\textbf {\bibinfo
  {volume} {509}},\ \bibinfo {pages} {166711} (\bibinfo {year}
  {2020})}\BibitemShut {NoStop}%
\bibitem [{\citenamefont {Jermain}\ \emph {et~al.}(2017)\citenamefont
  {Jermain}, \citenamefont {Aradhya}, \citenamefont {Reynolds}, \citenamefont
  {Buhrman}, \citenamefont {Brangham}, \citenamefont {Page}, \citenamefont
  {Hammel}, \citenamefont {Yang},\ and\ \citenamefont {Ralph}}]{Jermain2017}%
  \BibitemOpen
  \bibfield  {author} {\bibinfo {author} {\bibfnamefont {C.~L.}\ \bibnamefont
  {Jermain}}, \bibinfo {author} {\bibfnamefont {S.~V.}\ \bibnamefont
  {Aradhya}}, \bibinfo {author} {\bibfnamefont {N.~D.}\ \bibnamefont
  {Reynolds}}, \bibinfo {author} {\bibfnamefont {R.~A.}\ \bibnamefont
  {Buhrman}}, \bibinfo {author} {\bibfnamefont {J.~T.}\ \bibnamefont
  {Brangham}}, \bibinfo {author} {\bibfnamefont {M.~R.}\ \bibnamefont {Page}},
  \bibinfo {author} {\bibfnamefont {P.~C.}\ \bibnamefont {Hammel}}, \bibinfo
  {author} {\bibfnamefont {F.~Y.}\ \bibnamefont {Yang}},\ and\ \bibinfo
  {author} {\bibfnamefont {D.~C.}\ \bibnamefont {Ralph}},\ }\bibfield  {title}
  {\bibinfo {title} {Increased low-temperature damping in yttrium iron garnet
  thin films},\ }\href {https://doi.org/10.1103/PhysRevB.95.174411} {\bibfield
  {journal} {\bibinfo  {journal} {Phys. Rev. B}\ }\textbf {\bibinfo {volume}
  {95}},\ \bibinfo {pages} {174411} (\bibinfo {year} {2017})}\BibitemShut
  {NoStop}%
\bibitem [{\citenamefont {Danilov}\ \emph {et~al.}(1989)\citenamefont
  {Danilov}, \citenamefont {Lyfar'}, \citenamefont {Lyubon'ko}, \citenamefont
  {Nechiporuk},\ and\ \citenamefont {Ryabchenko}}]{Danilov1989}%
  \BibitemOpen
  \bibfield  {author} {\bibinfo {author} {\bibfnamefont {V.}~\bibnamefont
  {Danilov}}, \bibinfo {author} {\bibfnamefont {D.}~\bibnamefont {Lyfar'}},
  \bibinfo {author} {\bibfnamefont {Y.~V.}\ \bibnamefont {Lyubon'ko}}, \bibinfo
  {author} {\bibfnamefont {A.~Y.}\ \bibnamefont {Nechiporuk}},\ and\ \bibinfo
  {author} {\bibfnamefont {S.}~\bibnamefont {Ryabchenko}},\ }\bibfield  {title}
  {\bibinfo {title} {Low-temperature ferromagnetic resonance in epitaxial
  garnet films on paramagnetic substrates},\ }\href@noop {} {\bibfield
  {journal} {\bibinfo  {journal} {Soviet Physics Journal}\ }\textbf {\bibinfo
  {volume} {32}},\ \bibinfo {pages} {276} (\bibinfo {year} {1989})}\BibitemShut
  {NoStop}%
\bibitem [{\citenamefont {Danilov}\ and\ \citenamefont
  {Nechiporuk}(2002)}]{Danilov2002}%
  \BibitemOpen
  \bibfield  {author} {\bibinfo {author} {\bibfnamefont {V.}~\bibnamefont
  {Danilov}}\ and\ \bibinfo {author} {\bibfnamefont {A.~Y.}\ \bibnamefont
  {Nechiporuk}},\ }\bibfield  {title} {\bibinfo {title} {Experimental
  investigation of the quantum amplification effect for magnetostatic waves in
  ferrite-paramagnet structures},\ }\href@noop {} {\bibfield  {journal}
  {\bibinfo  {journal} {Technical Physics Letters}\ }\textbf {\bibinfo {volume}
  {28}},\ \bibinfo {pages} {369} (\bibinfo {year} {2002})}\BibitemShut
  {NoStop}%
\bibitem [{\citenamefont {Mihalceanu}\ \emph {et~al.}(2018)\citenamefont
  {Mihalceanu}, \citenamefont {Vasyuchka}, \citenamefont {Bozhko},
  \citenamefont {Langner}, \citenamefont {Nechiporuk}, \citenamefont
  {Romanyuk}, \citenamefont {Hillebrands},\ and\ \citenamefont
  {Serga}}]{Mihalcenu2018}%
  \BibitemOpen
  \bibfield  {author} {\bibinfo {author} {\bibfnamefont {L.}~\bibnamefont
  {Mihalceanu}}, \bibinfo {author} {\bibfnamefont {V.~I.}\ \bibnamefont
  {Vasyuchka}}, \bibinfo {author} {\bibfnamefont {D.~A.}\ \bibnamefont
  {Bozhko}}, \bibinfo {author} {\bibfnamefont {T.}~\bibnamefont {Langner}},
  \bibinfo {author} {\bibfnamefont {A.~Y.}\ \bibnamefont {Nechiporuk}},
  \bibinfo {author} {\bibfnamefont {V.~F.}\ \bibnamefont {Romanyuk}}, \bibinfo
  {author} {\bibfnamefont {B.}~\bibnamefont {Hillebrands}},\ and\ \bibinfo
  {author} {\bibfnamefont {A.~A.}\ \bibnamefont {Serga}},\ }\bibfield  {title}
  {\bibinfo {title} {Temperature-dependent relaxation of dipole-exchange
  magnons in yttrium iron garnet films},\ }\href
  {https://doi.org/10.1103/PhysRevB.97.214405} {\bibfield  {journal} {\bibinfo
  {journal} {Phys. Rev. B}\ }\textbf {\bibinfo {volume} {97}},\ \bibinfo
  {pages} {214405} (\bibinfo {year} {2018})}\BibitemShut {NoStop}%
\bibitem [{\citenamefont {Kosen}\ \emph {et~al.}(2019)\citenamefont {Kosen},
  \citenamefont {van Loo}, \citenamefont {Bozhko}, \citenamefont {Mihalceanu},\
  and\ \citenamefont {Karenowska}}]{Kosen2019}%
  \BibitemOpen
  \bibfield  {author} {\bibinfo {author} {\bibfnamefont {S.}~\bibnamefont
  {Kosen}}, \bibinfo {author} {\bibfnamefont {A.~F.}\ \bibnamefont {van Loo}},
  \bibinfo {author} {\bibfnamefont {D.~A.}\ \bibnamefont {Bozhko}}, \bibinfo
  {author} {\bibfnamefont {L.}~\bibnamefont {Mihalceanu}},\ and\ \bibinfo
  {author} {\bibfnamefont {A.~D.}\ \bibnamefont {Karenowska}},\ }\bibfield
  {title} {\bibinfo {title} {{Microwave magnon damping in YIG films at
  millikelvin temperatures}},\ }\href {https://doi.org/10.1063/1.5115266}
  {\bibfield  {journal} {\bibinfo  {journal} {APL Materials}\ }\textbf
  {\bibinfo {volume} {7}},\ \bibinfo {pages} {101120} (\bibinfo {year}
  {2019})},\ \Eprint
  {https://arxiv.org/abs/https://pubs.aip.org/aip/apm/article-pdf/doi/10.1063/1.5115266/14563526/101120\_1\_online.pdf}
  {https://pubs.aip.org/aip/apm/article-pdf/doi/10.1063/1.5115266/14563526/101120\_1\_online.pdf}
  \BibitemShut {NoStop}%
\bibitem [{\citenamefont {Azzawi}\ \emph {et~al.}(2023)\citenamefont {Azzawi},
  \citenamefont {Umerski}, \citenamefont {Sampaio}, \citenamefont {Bunyaev},
  \citenamefont {Kakazei},\ and\ \citenamefont {Atkinson}}]{Azzawi2023}%
  \BibitemOpen
  \bibfield  {author} {\bibinfo {author} {\bibfnamefont {S.}~\bibnamefont
  {Azzawi}}, \bibinfo {author} {\bibfnamefont {A.}~\bibnamefont {Umerski}},
  \bibinfo {author} {\bibfnamefont {L.~C.}\ \bibnamefont {Sampaio}}, \bibinfo
  {author} {\bibfnamefont {S.~A.}\ \bibnamefont {Bunyaev}}, \bibinfo {author}
  {\bibfnamefont {G.~N.}\ \bibnamefont {Kakazei}},\ and\ \bibinfo {author}
  {\bibfnamefont {D.}~\bibnamefont {Atkinson}},\ }\bibfield  {title} {\bibinfo
  {title} {Synthetic route to low damping in ferromagnetic thin-films},\
  }\bibfield  {journal} {\bibinfo  {journal} {APL Materials}\ }\textbf
  {\bibinfo {volume} {11}},\ \href {https://doi.org/10.1063/5.0147172/2905618}
  {10.1063/5.0147172/2905618} (\bibinfo {year} {2023})\BibitemShut {NoStop}%
\bibitem [{\citenamefont {Korenman}\ and\ \citenamefont
  {Prange}(1972)}]{Korenman1972}%
  \BibitemOpen
  \bibfield  {author} {\bibinfo {author} {\bibfnamefont {V.}~\bibnamefont
  {Korenman}}\ and\ \bibinfo {author} {\bibfnamefont {R.~E.}\ \bibnamefont
  {Prange}},\ }\bibfield  {title} {\bibinfo {title} {Anomalous damping of spin
  waves in magnetic metals},\ }\href {https://doi.org/10.1103/PhysRevB.6.2769}
  {\bibfield  {journal} {\bibinfo  {journal} {Physical Review B}\ }\textbf
  {\bibinfo {volume} {6}},\ \bibinfo {pages} {2769} (\bibinfo {year}
  {1972})}\BibitemShut {NoStop}%
\bibitem [{\citenamefont {Kamberský}(1976)}]{Kambersky1976}%
  \BibitemOpen
  \bibfield  {author} {\bibinfo {author} {\bibfnamefont {V.}~\bibnamefont
  {Kamberský}},\ }\bibfield  {title} {\bibinfo {title} {On ferromagnetic
  resonance damping in metals},\ }\href
  {https://doi.org/10.1007/BF01587621/METRICS} {\bibfield  {journal} {\bibinfo
  {journal} {Czechoslovak Journal of Physics}\ }\textbf {\bibinfo {volume}
  {26}},\ \bibinfo {pages} {1366} (\bibinfo {year} {1976})}\BibitemShut
  {NoStop}%
\bibitem [{\citenamefont {Gilmore}\ \emph {et~al.}(2007)\citenamefont
  {Gilmore}, \citenamefont {Idzerda},\ and\ \citenamefont
  {Stiles}}]{Gilmore2007}%
  \BibitemOpen
  \bibfield  {author} {\bibinfo {author} {\bibfnamefont {K.}~\bibnamefont
  {Gilmore}}, \bibinfo {author} {\bibfnamefont {Y.~U.}\ \bibnamefont
  {Idzerda}},\ and\ \bibinfo {author} {\bibfnamefont {M.~D.}\ \bibnamefont
  {Stiles}},\ }\bibfield  {title} {\bibinfo {title} {{Identification of the
  dominant precession-damping mechanism in Fe, Co, and Ni by first-principles
  calculations}},\ }\href {https://doi.org/10.1103/physrevlett.99.027204}
  {\bibfield  {journal} {\bibinfo  {journal} {Phys. Rev. Lett.}\ }\textbf
  {\bibinfo {volume} {99}},\ \bibinfo {pages} {027204} (\bibinfo {year}
  {2007})}\BibitemShut {NoStop}%
\bibitem [{\citenamefont {Kamberský}(1970)}]{Kambersky1970}%
  \BibitemOpen
  \bibfield  {author} {\bibinfo {author} {\bibfnamefont {V.}~\bibnamefont
  {Kamberský}},\ }\bibfield  {title} {\bibinfo {title} {{On the
  Landau–Lifshitz relaxation in ferromagnetic metals}},\ }\href
  {https://doi.org/10.1139/p70-361} {\bibfield  {journal} {\bibinfo  {journal}
  {Canadian Journal of Physics}\ }\textbf {\bibinfo {volume} {48}},\ \bibinfo
  {pages} {2906} (\bibinfo {year} {1970})}\BibitemShut {NoStop}%
\bibitem [{\citenamefont {Kamberský}(2007)}]{Kambersky2007}%
  \BibitemOpen
  \bibfield  {author} {\bibinfo {author} {\bibfnamefont {V.}~\bibnamefont
  {Kamberský}},\ }\bibfield  {title} {\bibinfo {title} {{Spin-orbital Gilbert
  damping in common magnetic metals}},\ }\href
  {https://doi.org/10.1103/PHYSREVB.76.134416/FIGURES/3/MEDIUM} {\bibfield
  {journal} {\bibinfo  {journal} {Physical Review B - Condensed Matter and
  Materials Physics}\ }\textbf {\bibinfo {volume} {76}},\ \bibinfo {pages}
  {134416} (\bibinfo {year} {2007})}\BibitemShut {NoStop}%
\bibitem [{\citenamefont {Gilmore}\ \emph {et~al.}(2008)\citenamefont
  {Gilmore}, \citenamefont {Idzerda},\ and\ \citenamefont
  {Stiles}}]{Gilmore2008}%
  \BibitemOpen
  \bibfield  {author} {\bibinfo {author} {\bibfnamefont {K.}~\bibnamefont
  {Gilmore}}, \bibinfo {author} {\bibfnamefont {Y.~U.}\ \bibnamefont
  {Idzerda}},\ and\ \bibinfo {author} {\bibfnamefont {M.~D.}\ \bibnamefont
  {Stiles}},\ }\bibfield  {title} {\bibinfo {title} {Spin-orbit precession
  damping in transition metal ferromagnets (invited)},\ }\href
  {https://doi.org/10.1063/1.2832348/343841} {\bibfield  {journal} {\bibinfo
  {journal} {Journal of Applied Physics}\ }\textbf {\bibinfo {volume} {103}},\
  \bibinfo {pages} {7} (\bibinfo {year} {2008})}\BibitemShut {NoStop}%
\bibitem [{\citenamefont {Thonig}\ and\ \citenamefont
  {Henk}(2014)}]{Thonig2014}%
  \BibitemOpen
  \bibfield  {author} {\bibinfo {author} {\bibfnamefont {D.}~\bibnamefont
  {Thonig}}\ and\ \bibinfo {author} {\bibfnamefont {J.}~\bibnamefont {Henk}},\
  }\bibfield  {title} {\bibinfo {title} {Gilbert damping tensor within the
  breathing fermi surface model: anisotropy and non-locality},\ }\href
  {https://doi.org/10.1088/1367-2630/16/1/013032} {\bibfield  {journal}
  {\bibinfo  {journal} {New Journal of Physics}\ }\textbf {\bibinfo {volume}
  {16}},\ \bibinfo {pages} {13032} (\bibinfo {year} {2014})}\BibitemShut
  {NoStop}%
\bibitem [{\citenamefont {Mizukami}\ \emph {et~al.}(2011)\citenamefont
  {Mizukami}, \citenamefont {Wu}, \citenamefont {Sakuma}, \citenamefont
  {Walowski}, \citenamefont {Watanabe}, \citenamefont {Kubota}, \citenamefont
  {Zhang}, \citenamefont {Naganuma}, \citenamefont {Oogane}, \citenamefont
  {Ando},\ and\ \citenamefont {Miyazaki}}]{Mizukami2011}%
  \BibitemOpen
  \bibfield  {author} {\bibinfo {author} {\bibfnamefont {S.}~\bibnamefont
  {Mizukami}}, \bibinfo {author} {\bibfnamefont {F.}~\bibnamefont {Wu}},
  \bibinfo {author} {\bibfnamefont {A.}~\bibnamefont {Sakuma}}, \bibinfo
  {author} {\bibfnamefont {J.}~\bibnamefont {Walowski}}, \bibinfo {author}
  {\bibfnamefont {D.}~\bibnamefont {Watanabe}}, \bibinfo {author}
  {\bibfnamefont {T.}~\bibnamefont {Kubota}}, \bibinfo {author} {\bibfnamefont
  {X.}~\bibnamefont {Zhang}}, \bibinfo {author} {\bibfnamefont
  {H.}~\bibnamefont {Naganuma}}, \bibinfo {author} {\bibfnamefont
  {M.}~\bibnamefont {Oogane}}, \bibinfo {author} {\bibfnamefont
  {Y.}~\bibnamefont {Ando}},\ and\ \bibinfo {author} {\bibfnamefont
  {T.}~\bibnamefont {Miyazaki}},\ }\bibfield  {title} {\bibinfo {title}
  {Long-lived ultrafast spin precession in manganese alloys films with a large
  perpendicular magnetic anisotropy},\ }\href
  {https://doi.org/10.1103/PHYSREVLETT.106.117201/FIGURES/3/MEDIUM} {\bibfield
  {journal} {\bibinfo  {journal} {Physical Review Letters}\ }\textbf {\bibinfo
  {volume} {106}},\ \bibinfo {pages} {117201} (\bibinfo {year}
  {2011})}\BibitemShut {NoStop}%
\bibitem [{\citenamefont {Schoen}\ \emph {et~al.}(2016)\citenamefont {Schoen},
  \citenamefont {Thonig}, \citenamefont {Schneider}, \citenamefont {Silva},
  \citenamefont {Nembach}, \citenamefont {Eriksson}, \citenamefont {Karis},\
  and\ \citenamefont {Shaw}}]{Schoen2016}%
  \BibitemOpen
  \bibfield  {author} {\bibinfo {author} {\bibfnamefont {M.~A.}\ \bibnamefont
  {Schoen}}, \bibinfo {author} {\bibfnamefont {D.}~\bibnamefont {Thonig}},
  \bibinfo {author} {\bibfnamefont {M.~L.}\ \bibnamefont {Schneider}}, \bibinfo
  {author} {\bibfnamefont {T.~J.}\ \bibnamefont {Silva}}, \bibinfo {author}
  {\bibfnamefont {H.~T.}\ \bibnamefont {Nembach}}, \bibinfo {author}
  {\bibfnamefont {O.}~\bibnamefont {Eriksson}}, \bibinfo {author}
  {\bibfnamefont {O.}~\bibnamefont {Karis}},\ and\ \bibinfo {author}
  {\bibfnamefont {J.~M.}\ \bibnamefont {Shaw}},\ }\bibfield  {title} {\bibinfo
  {title} {Ultra-low magnetic damping of a metallic ferromagnet},\ }\href
  {https://doi.org/10.1038/nphys3770} {\bibfield  {journal} {\bibinfo
  {journal} {Nature Physics 2016 12:9}\ }\textbf {\bibinfo {volume} {12}},\
  \bibinfo {pages} {839} (\bibinfo {year} {2016})}\BibitemShut {NoStop}%
\bibitem [{\citenamefont {Schoen}\ \emph {et~al.}(2017)\citenamefont {Schoen},
  \citenamefont {Lucassen}, \citenamefont {Nembach}, \citenamefont {Koopmans},
  \citenamefont {Silva}, \citenamefont {Back},\ and\ \citenamefont
  {Shaw}}]{Schoen2017}%
  \BibitemOpen
  \bibfield  {author} {\bibinfo {author} {\bibfnamefont {M.~A.}\ \bibnamefont
  {Schoen}}, \bibinfo {author} {\bibfnamefont {J.}~\bibnamefont {Lucassen}},
  \bibinfo {author} {\bibfnamefont {H.~T.}\ \bibnamefont {Nembach}}, \bibinfo
  {author} {\bibfnamefont {B.}~\bibnamefont {Koopmans}}, \bibinfo {author}
  {\bibfnamefont {T.~J.}\ \bibnamefont {Silva}}, \bibinfo {author}
  {\bibfnamefont {C.~H.}\ \bibnamefont {Back}},\ and\ \bibinfo {author}
  {\bibfnamefont {J.~M.}\ \bibnamefont {Shaw}},\ }\bibfield  {title} {\bibinfo
  {title} {{Magnetic properties in ultrathin 3d transition-metal binary alloys.
  II. Experimental verification of quantitative theories of damping and spin
  pumping}},\ }\href
  {https://doi.org/10.1103/PHYSREVB.95.134411/FIGURES/5/MEDIUM} {\bibfield
  {journal} {\bibinfo  {journal} {Physical Review B}\ }\textbf {\bibinfo
  {volume} {95}},\ \bibinfo {pages} {134411} (\bibinfo {year}
  {2017})}\BibitemShut {NoStop}%
\bibitem [{\citenamefont {Arora}\ \emph {et~al.}(2021)\citenamefont {Arora},
  \citenamefont {Delczeg-Czirjak}, \citenamefont {Riley}, \citenamefont
  {Silva}, \citenamefont {Nembach}, \citenamefont {Eriksson},\ and\
  \citenamefont {Shaw}}]{Arora2021}%
  \BibitemOpen
  \bibfield  {author} {\bibinfo {author} {\bibfnamefont {M.}~\bibnamefont
  {Arora}}, \bibinfo {author} {\bibfnamefont {E.~K.}\ \bibnamefont
  {Delczeg-Czirjak}}, \bibinfo {author} {\bibfnamefont {G.}~\bibnamefont
  {Riley}}, \bibinfo {author} {\bibfnamefont {T.~J.}\ \bibnamefont {Silva}},
  \bibinfo {author} {\bibfnamefont {H.~T.}\ \bibnamefont {Nembach}}, \bibinfo
  {author} {\bibfnamefont {O.}~\bibnamefont {Eriksson}},\ and\ \bibinfo
  {author} {\bibfnamefont {J.~M.}\ \bibnamefont {Shaw}},\ }\bibfield  {title}
  {\bibinfo {title} {{Magnetic Damping in Polycrystalline Thin-Film Fe-V
  Alloys}},\ }\href
  {https://doi.org/10.1103/PHYSREVAPPLIED.15.054031/FIGURES/8/MEDIUM}
  {\bibfield  {journal} {\bibinfo  {journal} {Physical Review Applied}\
  }\textbf {\bibinfo {volume} {15}},\ \bibinfo {pages} {054031} (\bibinfo
  {year} {2021})}\BibitemShut {NoStop}%
\bibitem [{\citenamefont {Wang}\ \emph {et~al.}(2019)\citenamefont {Wang},
  \citenamefont {Dong}, \citenamefont {Wei}, \citenamefont {Lin}, \citenamefont
  {Athey}, \citenamefont {Chen}, \citenamefont {Winter}, \citenamefont
  {Stephen}, \citenamefont {Heiman}, \citenamefont {He}, \citenamefont {Chen},
  \citenamefont {Liang}, \citenamefont {Yu}, \citenamefont {Zhang},
  \citenamefont {Podlaha-Murphy}, \citenamefont {Zhu}, \citenamefont {Wang},
  \citenamefont {Ni}, \citenamefont {McConney}, \citenamefont {Jones},
  \citenamefont {Page}, \citenamefont {Mahalingam},\ and\ \citenamefont
  {Sun}}]{Wang2019}%
  \BibitemOpen
  \bibfield  {author} {\bibinfo {author} {\bibfnamefont {J.}~\bibnamefont
  {Wang}}, \bibinfo {author} {\bibfnamefont {C.}~\bibnamefont {Dong}}, \bibinfo
  {author} {\bibfnamefont {Y.}~\bibnamefont {Wei}}, \bibinfo {author}
  {\bibfnamefont {X.}~\bibnamefont {Lin}}, \bibinfo {author} {\bibfnamefont
  {B.}~\bibnamefont {Athey}}, \bibinfo {author} {\bibfnamefont
  {Y.}~\bibnamefont {Chen}}, \bibinfo {author} {\bibfnamefont {A.}~\bibnamefont
  {Winter}}, \bibinfo {author} {\bibfnamefont {G.~M.}\ \bibnamefont {Stephen}},
  \bibinfo {author} {\bibfnamefont {D.}~\bibnamefont {Heiman}}, \bibinfo
  {author} {\bibfnamefont {Y.}~\bibnamefont {He}}, \bibinfo {author}
  {\bibfnamefont {H.}~\bibnamefont {Chen}}, \bibinfo {author} {\bibfnamefont
  {X.}~\bibnamefont {Liang}}, \bibinfo {author} {\bibfnamefont
  {C.}~\bibnamefont {Yu}}, \bibinfo {author} {\bibfnamefont {Y.}~\bibnamefont
  {Zhang}}, \bibinfo {author} {\bibfnamefont {E.~J.}\ \bibnamefont
  {Podlaha-Murphy}}, \bibinfo {author} {\bibfnamefont {M.}~\bibnamefont {Zhu}},
  \bibinfo {author} {\bibfnamefont {X.}~\bibnamefont {Wang}}, \bibinfo {author}
  {\bibfnamefont {J.}~\bibnamefont {Ni}}, \bibinfo {author} {\bibfnamefont
  {M.}~\bibnamefont {McConney}}, \bibinfo {author} {\bibfnamefont
  {J.}~\bibnamefont {Jones}}, \bibinfo {author} {\bibfnamefont
  {M.}~\bibnamefont {Page}}, \bibinfo {author} {\bibfnamefont {K.}~\bibnamefont
  {Mahalingam}},\ and\ \bibinfo {author} {\bibfnamefont {N.~X.}\ \bibnamefont
  {Sun}},\ }\bibfield  {title} {\bibinfo {title} {{Magnetostriction, Soft
  Magnetism, and Microwave Properties in Co-Fe-C Alloy Films}},\ }\href
  {https://doi.org/10.1103/PHYSREVAPPLIED.12.034011/FIGURES/6/MEDIUM}
  {\bibfield  {journal} {\bibinfo  {journal} {Physical Review Applied}\
  }\textbf {\bibinfo {volume} {12}},\ \bibinfo {pages} {034011} (\bibinfo
  {year} {2019})}\BibitemShut {NoStop}%
\bibitem [{\citenamefont {Lourembam}\ \emph {et~al.}(2021)\citenamefont
  {Lourembam}, \citenamefont {Khoo}, \citenamefont {Qiu}, \citenamefont {Xie},
  \citenamefont {Wong}, \citenamefont {Yap},\ and\ \citenamefont
  {Lim}}]{Lourembam2021}%
  \BibitemOpen
  \bibfield  {author} {\bibinfo {author} {\bibfnamefont {J.}~\bibnamefont
  {Lourembam}}, \bibinfo {author} {\bibfnamefont {K.~H.}\ \bibnamefont {Khoo}},
  \bibinfo {author} {\bibfnamefont {J.}~\bibnamefont {Qiu}}, \bibinfo {author}
  {\bibfnamefont {H.}~\bibnamefont {Xie}}, \bibinfo {author} {\bibfnamefont
  {S.~K.}\ \bibnamefont {Wong}}, \bibinfo {author} {\bibfnamefont {Q.~J.}\
  \bibnamefont {Yap}},\ and\ \bibinfo {author} {\bibfnamefont {S.~T.}\
  \bibnamefont {Lim}},\ }\bibfield  {title} {\bibinfo {title} {{Tuning Damping
  and Magnetic Anisotropy in Ultrathin Boron-Engineered MgO/Co--Fe--B/MgO
  Heterostructures}},\ }\href@noop {} {\bibfield  {journal} {\bibinfo
  {journal} {Advanced Electronic Materials}\ }\textbf {\bibinfo {volume} {7}},\
  \bibinfo {pages} {2100351} (\bibinfo {year} {2021})}\BibitemShut {NoStop}%
\bibitem [{\citenamefont {Momma}\ and\ \citenamefont
  {Izumi}(2011)}]{Momma2011}%
  \BibitemOpen
  \bibfield  {author} {\bibinfo {author} {\bibfnamefont {K.}~\bibnamefont
  {Momma}}\ and\ \bibinfo {author} {\bibfnamefont {F.}~\bibnamefont {Izumi}},\
  }\bibfield  {title} {\bibinfo {title} {Vesta 3 for three-dimensional
  visualization of crystal, volumetric and morphology data},\ }\href
  {https://doi.org/10.1107/S0021889811038970} {\bibfield  {journal} {\bibinfo
  {journal} {urn:issn:0021-8898}\ }\textbf {\bibinfo {volume} {44}},\ \bibinfo
  {pages} {1272} (\bibinfo {year} {2011})}\BibitemShut {NoStop}%
\bibitem [{\citenamefont {Åberg}\ \emph {et~al.}(2013)\citenamefont {Åberg},
  \citenamefont {Erhart},\ and\ \citenamefont {Lordi}}]{Aberg2013}%
  \BibitemOpen
  \bibfield  {author} {\bibinfo {author} {\bibfnamefont {D.}~\bibnamefont
  {Åberg}}, \bibinfo {author} {\bibfnamefont {P.}~\bibnamefont {Erhart}},\
  and\ \bibinfo {author} {\bibfnamefont {V.}~\bibnamefont {Lordi}},\ }\bibfield
   {title} {\bibinfo {title} {{Contributions of point defects, chemical
  disorder, and thermal vibrations to electronic properties of
  $\mathrm{Cd_{1-x}Zn_xTe}$ alloys}},\ }\href
  {https://doi.org/10.1103/PHYSREVB.88.045201/FIGURES/13/MEDIUM} {\bibfield
  {journal} {\bibinfo  {journal} {Physical Review B - Condensed Matter and
  Materials Physics}\ }\textbf {\bibinfo {volume} {88}},\ \bibinfo {pages}
  {045201} (\bibinfo {year} {2013})}\BibitemShut {NoStop}%
\bibitem [{\citenamefont {Schleife}\ \emph {et~al.}(2010)\citenamefont
  {Schleife}, \citenamefont {Eisenacher}, \citenamefont {Rödl}, \citenamefont
  {Fuchs}, \citenamefont {Furthmüller},\ and\ \citenamefont
  {Bechstedt}}]{Schleife2010}%
  \BibitemOpen
  \bibfield  {author} {\bibinfo {author} {\bibfnamefont {A.}~\bibnamefont
  {Schleife}}, \bibinfo {author} {\bibfnamefont {M.}~\bibnamefont
  {Eisenacher}}, \bibinfo {author} {\bibfnamefont {C.}~\bibnamefont {Rödl}},
  \bibinfo {author} {\bibfnamefont {F.}~\bibnamefont {Fuchs}}, \bibinfo
  {author} {\bibfnamefont {J.}~\bibnamefont {Furthmüller}},\ and\ \bibinfo
  {author} {\bibfnamefont {F.}~\bibnamefont {Bechstedt}},\ }\bibfield  {title}
  {\bibinfo {title} {Ab initio description of heterostructural alloys:
  Thermodynamic and structural properties of $\mathrm{Mg_xZn_{1-x}O}$ and
  $\mathrm{Cd_xZn_{1-x}O}$},\ }\href
  {https://journals.aps.org/prb/abstract/10.1103/PhysRevB.81.245210} {\bibfield
   {journal} {\bibinfo  {journal} {Physical Review B}\ }\textbf {\bibinfo
  {volume} {81}},\ \bibinfo {pages} {245210} (\bibinfo {year}
  {2010})}\BibitemShut {NoStop}%
\bibitem [{\citenamefont {Schleife}\ \emph {et~al.}(2011)\citenamefont
  {Schleife}, \citenamefont {R{\"o}dl}, \citenamefont {Furthm{\"u}ller},\ and\
  \citenamefont {Bechstedt}}]{Schleife2011}%
  \BibitemOpen
  \bibfield  {author} {\bibinfo {author} {\bibfnamefont {A.}~\bibnamefont
  {Schleife}}, \bibinfo {author} {\bibfnamefont {C.}~\bibnamefont {R{\"o}dl}},
  \bibinfo {author} {\bibfnamefont {J.}~\bibnamefont {Furthm{\"u}ller}},\ and\
  \bibinfo {author} {\bibfnamefont {F.}~\bibnamefont {Bechstedt}},\ }\bibfield
  {title} {\bibinfo {title} {Electronic and optical properties of
  $\mathrm{Mg_xZn_{1-x}O}$ and $\mathrm{Cd_xZn_{1-x}O}$ from ab initio
  calculations},\ }\href
  {https://iopscience.iop.org/article/10.1088/1367-2630/13/8/085012} {\bibfield
   {journal} {\bibinfo  {journal} {New Journal of Physics}\ }\textbf {\bibinfo
  {volume} {13}},\ \bibinfo {pages} {085012} (\bibinfo {year}
  {2011})}\BibitemShut {NoStop}%
\bibitem [{\citenamefont {Sanchez}\ \emph {et~al.}(1984)\citenamefont
  {Sanchez}, \citenamefont {Ducastelle},\ and\ \citenamefont
  {Gratias}}]{Sanchez1984}%
  \BibitemOpen
  \bibfield  {author} {\bibinfo {author} {\bibfnamefont {J.~M.}\ \bibnamefont
  {Sanchez}}, \bibinfo {author} {\bibfnamefont {F.}~\bibnamefont
  {Ducastelle}},\ and\ \bibinfo {author} {\bibfnamefont {D.}~\bibnamefont
  {Gratias}},\ }\bibfield  {title} {\bibinfo {title} {Generalized cluster
  description of multicomponent systems},\ }\href
  {https://doi.org/10.1016/0378-4371(84)90096-7} {\bibfield  {journal}
  {\bibinfo  {journal} {Physica A: Statistical Mechanics and its Applications}\
  }\textbf {\bibinfo {volume} {128}},\ \bibinfo {pages} {334} (\bibinfo {year}
  {1984})}\BibitemShut {NoStop}%
\bibitem [{\citenamefont {Zunger}\ \emph {et~al.}(1994)\citenamefont {Zunger},
  \citenamefont {Turchi},\ and\ \citenamefont {Gonis}}]{Zunger1994}%
  \BibitemOpen
  \bibfield  {author} {\bibinfo {author} {\bibfnamefont {A.}~\bibnamefont
  {Zunger}}, \bibinfo {author} {\bibfnamefont {P.}~\bibnamefont {Turchi}},\
  and\ \bibinfo {author} {\bibfnamefont {A.}~\bibnamefont {Gonis}},\ }\bibfield
   {title} {\bibinfo {title} {Statics and dynamics of alloy phase
  transformations},\ }\href@noop {} {\bibfield  {journal} {\bibinfo  {journal}
  {NATO ASI Series. Series B, Physics}\ }\textbf {\bibinfo {volume} {319}}
  (\bibinfo {year} {1994})}\BibitemShut {NoStop}%
\bibitem [{\citenamefont {Teles}\ \emph {et~al.}(2000)\citenamefont {Teles},
  \citenamefont {Furthm{\"u}ller}, \citenamefont {Scolfaro}, \citenamefont
  {Leite},\ and\ \citenamefont {Bechstedt}}]{Teles2000}%
  \BibitemOpen
  \bibfield  {author} {\bibinfo {author} {\bibfnamefont {L.}~\bibnamefont
  {Teles}}, \bibinfo {author} {\bibfnamefont {J.}~\bibnamefont
  {Furthm{\"u}ller}}, \bibinfo {author} {\bibfnamefont {L.~M.~R.}\ \bibnamefont
  {Scolfaro}}, \bibinfo {author} {\bibfnamefont {J.}~\bibnamefont {Leite}},\
  and\ \bibinfo {author} {\bibfnamefont {F.}~\bibnamefont {Bechstedt}},\
  }\bibfield  {title} {\bibinfo {title} {First-principles calculations of the
  thermodynamic and structural properties of strained $\mathrm{In_xGa_{1-x}N}$
  and $\mathrm{Al_xGa_{1-x}N}$ alloys},\ }\href
  {https://doi.org/10.1103/PhysRevB.62.2475} {\bibfield  {journal} {\bibinfo
  {journal} {Physical Review B}\ }\textbf {\bibinfo {volume} {62}},\ \bibinfo
  {pages} {2475} (\bibinfo {year} {2000})}\BibitemShut {NoStop}%
\bibitem [{\citenamefont {Caetano}\ \emph {et~al.}(2006)\citenamefont
  {Caetano}, \citenamefont {Teles}, \citenamefont {Marques}, \citenamefont
  {Pino},\ and\ \citenamefont {Ferreira}}]{Caetano2006}%
  \BibitemOpen
  \bibfield  {author} {\bibinfo {author} {\bibfnamefont {C.}~\bibnamefont
  {Caetano}}, \bibinfo {author} {\bibfnamefont {L.~K.}\ \bibnamefont {Teles}},
  \bibinfo {author} {\bibfnamefont {M.}~\bibnamefont {Marques}}, \bibinfo
  {author} {\bibfnamefont {A.~D.}\ \bibnamefont {Pino}},\ and\ \bibinfo
  {author} {\bibfnamefont {L.~G.}\ \bibnamefont {Ferreira}},\ }\bibfield
  {title} {\bibinfo {title} {Phase stability, chemical bonds, and gap bowing of
  $\mathrm{In_xGa_{1-x}N}$ alloys: Comparison between cubic and wurtzite
  structures},\ }\href
  {https://doi.org/10.1103/PHYSREVB.74.045215/FIGURES/7/MEDIUM} {\bibfield
  {journal} {\bibinfo  {journal} {Physical Review B - Condensed Matter and
  Materials Physics}\ }\textbf {\bibinfo {volume} {74}},\ \bibinfo {pages}
  {045215} (\bibinfo {year} {2006})}\BibitemShut {NoStop}%
\bibitem [{\citenamefont {Thompson}\ \emph {et~al.}(2022)\citenamefont
  {Thompson}, \citenamefont {Aktulga}, \citenamefont {Berger}, \citenamefont
  {Bolintineanu}, \citenamefont {Brown}, \citenamefont {Crozier}, \citenamefont
  {in~'t Veld}, \citenamefont {Kohlmeyer}, \citenamefont {Moore}, \citenamefont
  {Nguyen}, \citenamefont {Shan}, \citenamefont {Stevens}, \citenamefont
  {Tranchida}, \citenamefont {Trott},\ and\ \citenamefont
  {Plimpton}}]{Thompson2022}%
  \BibitemOpen
  \bibfield  {author} {\bibinfo {author} {\bibfnamefont {A.~P.}\ \bibnamefont
  {Thompson}}, \bibinfo {author} {\bibfnamefont {H.~M.}\ \bibnamefont
  {Aktulga}}, \bibinfo {author} {\bibfnamefont {R.}~\bibnamefont {Berger}},
  \bibinfo {author} {\bibfnamefont {D.~S.}\ \bibnamefont {Bolintineanu}},
  \bibinfo {author} {\bibfnamefont {W.~M.}\ \bibnamefont {Brown}}, \bibinfo
  {author} {\bibfnamefont {P.~S.}\ \bibnamefont {Crozier}}, \bibinfo {author}
  {\bibfnamefont {P.~J.}\ \bibnamefont {in~'t Veld}}, \bibinfo {author}
  {\bibfnamefont {A.}~\bibnamefont {Kohlmeyer}}, \bibinfo {author}
  {\bibfnamefont {S.~G.}\ \bibnamefont {Moore}}, \bibinfo {author}
  {\bibfnamefont {T.~D.}\ \bibnamefont {Nguyen}}, \bibinfo {author}
  {\bibfnamefont {R.}~\bibnamefont {Shan}}, \bibinfo {author} {\bibfnamefont
  {M.~J.}\ \bibnamefont {Stevens}}, \bibinfo {author} {\bibfnamefont
  {J.}~\bibnamefont {Tranchida}}, \bibinfo {author} {\bibfnamefont
  {C.}~\bibnamefont {Trott}},\ and\ \bibinfo {author} {\bibfnamefont {S.~J.}\
  \bibnamefont {Plimpton}},\ }\bibfield  {title} {\bibinfo {title} {Lammps - a
  flexible simulation tool for particle-based materials modeling at the atomic,
  meso, and continuum scales},\ }\href
  {https://doi.org/10.1016/J.CPC.2021.108171} {\bibfield  {journal} {\bibinfo
  {journal} {Computer Physics Communications}\ }\textbf {\bibinfo {volume}
  {271}},\ \bibinfo {pages} {108171} (\bibinfo {year} {2022})}\BibitemShut
  {NoStop}%
\bibitem [{\citenamefont {Dudarev}\ and\ \citenamefont
  {Derlet}(2005)}]{Dudarev2005}%
  \BibitemOpen
  \bibfield  {author} {\bibinfo {author} {\bibfnamefont {S.~L.}\ \bibnamefont
  {Dudarev}}\ and\ \bibinfo {author} {\bibfnamefont {P.~M.}\ \bibnamefont
  {Derlet}},\ }\bibfield  {title} {\bibinfo {title} {A ‘magnetic’
  interatomic potential for molecular dynamics simulations},\ }\href
  {https://doi.org/10.1088/0953-8984/17/44/003} {\bibfield  {journal} {\bibinfo
   {journal} {Journal of Physics: Condensed Matter}\ }\textbf {\bibinfo
  {volume} {17}},\ \bibinfo {pages} {7097} (\bibinfo {year}
  {2005})}\BibitemShut {NoStop}%
\bibitem [{\citenamefont {L}\ and\ \citenamefont
  {M}(2007)}]{Dudarev2007correction}%
  \BibitemOpen
  \bibfield  {author} {\bibinfo {author} {\bibfnamefont {D.~S.}\ \bibnamefont
  {L}}\ and\ \bibinfo {author} {\bibfnamefont {D.~P.}\ \bibnamefont {M}},\
  }\bibfield  {title} {\bibinfo {title} {A `magnetic' interatomic potential for
  molecular dynamics simulations.},\ }\href
  {https://doi.org/10.1088/0953-8984/19/23/239001} {\bibfield  {journal}
  {\bibinfo  {journal} {Journal of Physics: Condensed Matter}\ }\textbf
  {\bibinfo {volume} {19}},\ \bibinfo {pages} {239001} (\bibinfo {year}
  {2007})}\BibitemShut {NoStop}%
\bibitem [{\citenamefont {Derlet}\ and\ \citenamefont
  {Dudarev}(2007)}]{Derlet2007}%
  \BibitemOpen
  \bibfield  {author} {\bibinfo {author} {\bibfnamefont {P.~M.}\ \bibnamefont
  {Derlet}}\ and\ \bibinfo {author} {\bibfnamefont {S.~L.}\ \bibnamefont
  {Dudarev}},\ }\bibfield  {title} {\bibinfo {title} {Million-atom molecular
  dynamics simulations of magnetic iron},\ }\href
  {https://doi.org/10.1016/J.PMATSCI.2006.10.011} {\bibfield  {journal}
  {\bibinfo  {journal} {Progress in Materials Science}\ }\textbf {\bibinfo
  {volume} {52}},\ \bibinfo {pages} {299} (\bibinfo {year} {2007})}\BibitemShut
  {NoStop}%
\bibitem [{\citenamefont {Elliott}\ and\ \citenamefont
  {Tadmor}(2011)}]{OpenKim1}%
  \BibitemOpen
  \bibfield  {author} {\bibinfo {author} {\bibfnamefont {R.~S.}\ \bibnamefont
  {Elliott}}\ and\ \bibinfo {author} {\bibfnamefont {E.~B.}\ \bibnamefont
  {Tadmor}},\ }\href {https://doi.org/10.25950/ff8f563a} {\bibinfo {title}
  {{K}nowledgebase of {I}nteratomic {M}odels ({KIM}) application programming
  interface ({API})}},\ \bibinfo {howpublished}
  {\url{https://openkim.org/kim-api}} (\bibinfo {year} {2011})\BibitemShut
  {NoStop}%
\bibitem [{\citenamefont {Tadmor}\ \emph {et~al.}(2011)\citenamefont {Tadmor},
  \citenamefont {Elliott}, \citenamefont {Sethna}, \citenamefont {Miller},\
  and\ \citenamefont {Becker}}]{OpenKim2}%
  \BibitemOpen
  \bibfield  {author} {\bibinfo {author} {\bibfnamefont {E.~B.}\ \bibnamefont
  {Tadmor}}, \bibinfo {author} {\bibfnamefont {R.~S.}\ \bibnamefont {Elliott}},
  \bibinfo {author} {\bibfnamefont {J.~P.}\ \bibnamefont {Sethna}}, \bibinfo
  {author} {\bibfnamefont {R.~E.}\ \bibnamefont {Miller}},\ and\ \bibinfo
  {author} {\bibfnamefont {C.~A.}\ \bibnamefont {Becker}},\ }\bibfield  {title}
  {\bibinfo {title} {The potential of atomistic simulations and the
  {K}nowledgebase of {I}nteratomic {M}odels},\ }\href
  {https://doi.org/10.1007/s11837-011-0102-6} {\bibfield  {journal} {\bibinfo
  {journal} {{JOM}}\ }\textbf {\bibinfo {volume} {63}},\ \bibinfo {pages} {17}
  (\bibinfo {year} {2011})}\BibitemShut {NoStop}%
\bibitem [{\citenamefont {Gilbert}(2018)}]{OpenKim3}%
  \BibitemOpen
  \bibfield  {author} {\bibinfo {author} {\bibfnamefont {M.~R.}\ \bibnamefont
  {Gilbert}},\ }\href {https://doi.org/10.25950/9776664f} {\bibinfo {title}
  {{EAM} potential for magnetic bcc metals with cubic spline interpolation
  v002}},\ \bibinfo {howpublished} {OpenKIM,
  \url{https://doi.org/10.25950/9776664f}} (\bibinfo {year} {2018})\BibitemShut
  {NoStop}%
\bibitem [{\citenamefont {Dudarev}\ and\ \citenamefont
  {Derlet}(2018)}]{OpenKim4}%
  \BibitemOpen
  \bibfield  {author} {\bibinfo {author} {\bibfnamefont {S.}~\bibnamefont
  {Dudarev}}\ and\ \bibinfo {author} {\bibfnamefont {P.}~\bibnamefont
  {Derlet}},\ }\href {https://doi.org/10.25950/eb4996de} {\bibinfo {title}
  {{EAM} potential (magnetic, cubic tabulation) for magnetic {F}e developed by
  {D}udarev and {D}erlet (2005) v002}},\ \bibinfo {howpublished} {OpenKIM,
  \url{https://doi.org/10.25950/eb4996de}} (\bibinfo {year} {2018})\BibitemShut
  {NoStop}%
\bibitem [{\citenamefont {Ma}\ \emph {et~al.}(2007)\citenamefont {Ma},
  \citenamefont {Liu}, \citenamefont {Woo},\ and\ \citenamefont
  {Dudarev}}]{Ma2007}%
  \BibitemOpen
  \bibfield  {author} {\bibinfo {author} {\bibfnamefont {P.~W.}\ \bibnamefont
  {Ma}}, \bibinfo {author} {\bibfnamefont {W.~C.}\ \bibnamefont {Liu}},
  \bibinfo {author} {\bibfnamefont {C.~H.}\ \bibnamefont {Woo}},\ and\ \bibinfo
  {author} {\bibfnamefont {S.~L.}\ \bibnamefont {Dudarev}},\ }\bibfield
  {title} {\bibinfo {title} {Large-scale molecular dynamics simulation of
  magnetic properties of amorphous iron under pressure},\ }\href
  {https://doi.org/10.1063/1.2715753/919151} {\bibfield  {journal} {\bibinfo
  {journal} {Journal of Applied Physics}\ }\textbf {\bibinfo {volume} {101}},\
  \bibinfo {pages} {73908} (\bibinfo {year} {2007})}\BibitemShut {NoStop}%
\bibitem [{\citenamefont {Ichikawa}(1973)}]{Ichikawa1973}%
  \BibitemOpen
  \bibfield  {author} {\bibinfo {author} {\bibfnamefont {T.}~\bibnamefont
  {Ichikawa}},\ }\bibfield  {title} {\bibinfo {title} {Electron diffraction
  study of the local atomic arrangement in amorphous iron and nickel films},\
  }\href {https://doi.org/10.1002/PSSA.2210190237} {\bibfield  {journal}
  {\bibinfo  {journal} {physica status solidi (a)}\ }\textbf {\bibinfo {volume}
  {19}},\ \bibinfo {pages} {707} (\bibinfo {year} {1973})}\BibitemShut
  {NoStop}%
\bibitem [{\citenamefont {Kresse}\ and\ \citenamefont
  {Furthmüller}(1996{\natexlab{a}})}]{Kresse1996CMS}%
  \BibitemOpen
  \bibfield  {author} {\bibinfo {author} {\bibfnamefont {G.}~\bibnamefont
  {Kresse}}\ and\ \bibinfo {author} {\bibfnamefont {J.}~\bibnamefont
  {Furthmüller}},\ }\bibfield  {title} {\bibinfo {title} {Efficiency of
  ab-initio total energy calculations for metals and semiconductors using a
  plane-wave basis set},\ }\href {https://doi.org/10.1016/0927-0256(96)00008-0}
  {\bibfield  {journal} {\bibinfo  {journal} {Computational Materials Science}\
  }\textbf {\bibinfo {volume} {6}},\ \bibinfo {pages} {15} (\bibinfo {year}
  {1996}{\natexlab{a}})}\BibitemShut {NoStop}%
\bibitem [{\citenamefont {Kresse}\ and\ \citenamefont
  {Furthmüller}(1996{\natexlab{b}})}]{Kresse1996PRB}%
  \BibitemOpen
  \bibfield  {author} {\bibinfo {author} {\bibfnamefont {G.}~\bibnamefont
  {Kresse}}\ and\ \bibinfo {author} {\bibfnamefont {J.}~\bibnamefont
  {Furthmüller}},\ }\bibfield  {title} {\bibinfo {title} {Efficient iterative
  schemes for ab initio total-energy calculations using a plane-wave basis
  set},\ }\href {https://doi.org/10.1103/PhysRevB.54.11169} {\bibfield
  {journal} {\bibinfo  {journal} {Physical Review B}\ }\textbf {\bibinfo
  {volume} {54}},\ \bibinfo {pages} {11169} (\bibinfo {year}
  {1996}{\natexlab{b}})}\BibitemShut {NoStop}%
\bibitem [{\citenamefont {Perdew}\ \emph {et~al.}(1996)\citenamefont {Perdew},
  \citenamefont {Burke},\ and\ \citenamefont {Ernzerhof}}]{Perdew1996}%
  \BibitemOpen
  \bibfield  {author} {\bibinfo {author} {\bibfnamefont {J.~P.}\ \bibnamefont
  {Perdew}}, \bibinfo {author} {\bibfnamefont {K.}~\bibnamefont {Burke}},\ and\
  \bibinfo {author} {\bibfnamefont {M.}~\bibnamefont {Ernzerhof}},\ }\bibfield
  {title} {\bibinfo {title} {{Generalized Gradient Approximation Made
  Simple}},\ }\href {https://doi.org/10.1103/PhysRevLett.77.3865} {\bibfield
  {journal} {\bibinfo  {journal} {Physical Review Letters}\ }\textbf {\bibinfo
  {volume} {77}},\ \bibinfo {pages} {3865} (\bibinfo {year}
  {1996})}\BibitemShut {NoStop}%
\bibitem [{\citenamefont {Monkhorst}\ and\ \citenamefont
  {Pack}(1976)}]{Monkhorst1976}%
  \BibitemOpen
  \bibfield  {author} {\bibinfo {author} {\bibfnamefont {H.~J.}\ \bibnamefont
  {Monkhorst}}\ and\ \bibinfo {author} {\bibfnamefont {J.~D.}\ \bibnamefont
  {Pack}},\ }\bibfield  {title} {\bibinfo {title} {Special points for
  brillouin-zone integrations},\ }\href
  {https://doi.org/10.1103/PhysRevB.13.5188} {\bibfield  {journal} {\bibinfo
  {journal} {Physical Review B}\ }\textbf {\bibinfo {volume} {13}},\ \bibinfo
  {pages} {5188} (\bibinfo {year} {1976})}\BibitemShut {NoStop}%
\bibitem [{\citenamefont {Li}\ \emph {et~al.}(2019{\natexlab{b}})\citenamefont
  {Li}, \citenamefont {Zeng}, \citenamefont {Zhang}, \citenamefont {Shin},
  \citenamefont {Saglam}, \citenamefont {Karakas}, \citenamefont {Ozatay},
  \citenamefont {Pearson}, \citenamefont {Heinonen}, \citenamefont {Wu},
  \citenamefont {Hoffmann},\ and\ \citenamefont {Zhang}}]{Li2019AnisoDP}%
  \BibitemOpen
  \bibfield  {author} {\bibinfo {author} {\bibfnamefont {Y.}~\bibnamefont
  {Li}}, \bibinfo {author} {\bibfnamefont {F.}~\bibnamefont {Zeng}}, \bibinfo
  {author} {\bibfnamefont {S.~S.}\ \bibnamefont {Zhang}}, \bibinfo {author}
  {\bibfnamefont {H.}~\bibnamefont {Shin}}, \bibinfo {author} {\bibfnamefont
  {H.}~\bibnamefont {Saglam}}, \bibinfo {author} {\bibfnamefont
  {V.}~\bibnamefont {Karakas}}, \bibinfo {author} {\bibfnamefont
  {O.}~\bibnamefont {Ozatay}}, \bibinfo {author} {\bibfnamefont {J.~E.}\
  \bibnamefont {Pearson}}, \bibinfo {author} {\bibfnamefont {O.~G.}\
  \bibnamefont {Heinonen}}, \bibinfo {author} {\bibfnamefont {Y.}~\bibnamefont
  {Wu}}, \bibinfo {author} {\bibfnamefont {A.}~\bibnamefont {Hoffmann}},\ and\
  \bibinfo {author} {\bibfnamefont {W.}~\bibnamefont {Zhang}},\ }\bibfield
  {title} {\bibinfo {title} {{Giant Anisotropy of Gilbert Damping in Epitaxial
  CoFe Films}},\ }\href
  {https://doi.org/10.1103/PHYSREVLETT.122.117203/FIGURES/5/MEDIUM} {\bibfield
  {journal} {\bibinfo  {journal} {Physical Review Letters}\ }\textbf {\bibinfo
  {volume} {122}},\ \bibinfo {pages} {117203} (\bibinfo {year}
  {2019}{\natexlab{b}})}\BibitemShut {NoStop}%
\bibitem [{\citenamefont {Fähnle}\ \emph {et~al.}(2008)\citenamefont
  {Fähnle}, \citenamefont {Steiauf},\ and\ \citenamefont {Seib}}]{Fahnle2008}%
  \BibitemOpen
  \bibfield  {author} {\bibinfo {author} {\bibfnamefont {M.}~\bibnamefont
  {Fähnle}}, \bibinfo {author} {\bibfnamefont {D.}~\bibnamefont {Steiauf}},\
  and\ \bibinfo {author} {\bibfnamefont {J.}~\bibnamefont {Seib}},\ }\bibfield
  {title} {\bibinfo {title} {{The Gilbert equation revisited: anisotropic and
  nonlocal damping of magnetization dynamics}},\ }\href
  {https://doi.org/10.1088/0022-3727/41/16/164014} {\bibfield  {journal}
  {\bibinfo  {journal} {Journal of Physics D: Applied Physics}\ }\textbf
  {\bibinfo {volume} {41}},\ \bibinfo {pages} {164014} (\bibinfo {year}
  {2008})}\BibitemShut {NoStop}%
\bibitem [{\citenamefont {Khodadadi}\ \emph {et~al.}(2020)\citenamefont
  {Khodadadi}, \citenamefont {Rai}, \citenamefont {Sapkota}, \citenamefont
  {Srivastava}, \citenamefont {Nepal}, \citenamefont {Lim}, \citenamefont
  {Smith}, \citenamefont {Mewes}, \citenamefont {Budhathoki}, \citenamefont
  {Hauser}, \citenamefont {Gao}, \citenamefont {Li}, \citenamefont {Viehlad},
  \citenamefont {Jiag}, \citenamefont {Heremas}, \citenamefont {Balachadra},
  \citenamefont {Mewes},\ and\ \citenamefont {Emori}}]{Khodadadi2020}%
  \BibitemOpen
  \bibfield  {author} {\bibinfo {author} {\bibfnamefont {B.}~\bibnamefont
  {Khodadadi}}, \bibinfo {author} {\bibfnamefont {A.}~\bibnamefont {Rai}},
  \bibinfo {author} {\bibfnamefont {A.}~\bibnamefont {Sapkota}}, \bibinfo
  {author} {\bibfnamefont {A.}~\bibnamefont {Srivastava}}, \bibinfo {author}
  {\bibfnamefont {B.}~\bibnamefont {Nepal}}, \bibinfo {author} {\bibfnamefont
  {Y.}~\bibnamefont {Lim}}, \bibinfo {author} {\bibfnamefont {D.~A.}\
  \bibnamefont {Smith}}, \bibinfo {author} {\bibfnamefont {C.}~\bibnamefont
  {Mewes}}, \bibinfo {author} {\bibfnamefont {S.}~\bibnamefont {Budhathoki}},
  \bibinfo {author} {\bibfnamefont {A.~J.}\ \bibnamefont {Hauser}}, \bibinfo
  {author} {\bibfnamefont {M.}~\bibnamefont {Gao}}, \bibinfo {author}
  {\bibfnamefont {J.~F.}\ \bibnamefont {Li}}, \bibinfo {author} {\bibfnamefont
  {D.~D.}\ \bibnamefont {Viehlad}}, \bibinfo {author} {\bibfnamefont
  {Z.}~\bibnamefont {Jiag}}, \bibinfo {author} {\bibfnamefont {J.~J.}\
  \bibnamefont {Heremas}}, \bibinfo {author} {\bibfnamefont {P.~V.}\
  \bibnamefont {Balachadra}}, \bibinfo {author} {\bibfnamefont
  {T.}~\bibnamefont {Mewes}},\ and\ \bibinfo {author} {\bibfnamefont
  {S.}~\bibnamefont {Emori}},\ }\bibfield  {title} {\bibinfo {title}
  {{Conductivitylike Gilbert Damping due to Intraband Scattering in Epitaxial
  Iron}},\ }\href
  {https://doi.org/10.1103/PHYSREVLETT.124.157201/FIGURES/4/MEDIUM} {\bibfield
  {journal} {\bibinfo  {journal} {Physical Review Letters}\ }\textbf {\bibinfo
  {volume} {124}},\ \bibinfo {pages} {157201} (\bibinfo {year}
  {2020})}\BibitemShut {NoStop}%
\bibitem [{\citenamefont {Bhagat}\ and\ \citenamefont
  {Lubitz}(1974)}]{Bhagat1974}%
  \BibitemOpen
  \bibfield  {author} {\bibinfo {author} {\bibfnamefont {S.~M.}\ \bibnamefont
  {Bhagat}}\ and\ \bibinfo {author} {\bibfnamefont {P.}~\bibnamefont
  {Lubitz}},\ }\bibfield  {title} {\bibinfo {title} {Temperature variation of
  ferromagnetic relaxation in the 3d transition metals},\ }\href
  {https://doi.org/10.1103/PhysRevB.10.179} {\bibfield  {journal} {\bibinfo
  {journal} {Physical Review B}\ }\textbf {\bibinfo {volume} {10}},\ \bibinfo
  {pages} {179} (\bibinfo {year} {1974})}\BibitemShut {NoStop}%
\bibitem [{\citenamefont {Berry}\ and\ \citenamefont
  {Pritchet}(1975)}]{Berry1975}%
  \BibitemOpen
  \bibfield  {author} {\bibinfo {author} {\bibfnamefont {B.~S.}\ \bibnamefont
  {Berry}}\ and\ \bibinfo {author} {\bibfnamefont {W.~C.}\ \bibnamefont
  {Pritchet}},\ }\bibfield  {title} {\bibinfo {title} {Magnetic annealing and
  directional ordering of an amorphous ferromagnetic alloy},\ }\href
  {https://doi.org/10.1103/PhysRevLett.34.1022} {\bibfield  {journal} {\bibinfo
   {journal} {Phys. Rev. Lett.}\ }\textbf {\bibinfo {volume} {34}},\ \bibinfo
  {pages} {1022} (\bibinfo {year} {1975})}\BibitemShut {NoStop}%
\bibitem [{\citenamefont {Wu}\ \emph {et~al.}(2016)\citenamefont {Wu},
  \citenamefont {Wang}, \citenamefont {Li}, \citenamefont {Lou}, \citenamefont
  {Zhao}, \citenamefont {Gao},\ and\ \citenamefont {Wang}}]{Wu2016}%
  \BibitemOpen
  \bibfield  {author} {\bibinfo {author} {\bibfnamefont {C.}~\bibnamefont
  {Wu}}, \bibinfo {author} {\bibfnamefont {K.}~\bibnamefont {Wang}}, \bibinfo
  {author} {\bibfnamefont {D.}~\bibnamefont {Li}}, \bibinfo {author}
  {\bibfnamefont {C.}~\bibnamefont {Lou}}, \bibinfo {author} {\bibfnamefont
  {Y.}~\bibnamefont {Zhao}}, \bibinfo {author} {\bibfnamefont {Y.}~\bibnamefont
  {Gao}},\ and\ \bibinfo {author} {\bibfnamefont {Q.}~\bibnamefont {Wang}},\
  }\bibfield  {title} {\bibinfo {title} {{Tuning microstructure and magnetic
  properties of electrodeposited CoNiP films by high magnetic field
  annealing}},\ }\href@noop {} {\bibfield  {journal} {\bibinfo  {journal}
  {Journal of Magnetism and Magnetic Materials}\ }\textbf {\bibinfo {volume}
  {416}},\ \bibinfo {pages} {61} (\bibinfo {year} {2016})}\BibitemShut
  {NoStop}%
\end{thebibliography}%

\end{document}